\documentclass[twocolumn]{aastex62}
\usepackage{lineno}
\usepackage{multirow}
\usepackage{amsmath}
\usepackage{amsfonts}

\usepackage{todonotes}

\graphicspath{{./}{}}

\received{January 31, 2022}
\revised{March 20, 2022}
\accepted{\today}

\submitjournal{ApJ}

\shorttitle{Separating Mergers and Star Forming Galaxies using Deep Learning}
\shortauthors{Ferreira et al.}

\begin{document}

\title{A Simulation Driven Deep Learning Approach for Separating Mergers and Star Forming Galaxies: The Formation Histories of Clumpy Galaxies in all the CANDELS Fields}

\correspondingauthor{Leonardo Ferreira}
\email{leonardo.ferreira@nottingham.ac.uk, conselice@gmail.com}

\author[0000-0002-8919-079X]{Leonardo Ferreira}
\affil{Centre for Astronomy and Particle Physics, University of Nottingham, Nottingham, UK}

\author{Christopher J. Conselice}
\affil{Centre for Astronomy and Particle Physics, University of Nottingham, Nottingham, UK}
\affil{Jodrell Bank Centre for Astrophysics, University of Manchester, Oxford Road, Manchester UK}
\affil{Alan Turing Fellow, Alan Turing Institute, 96 Euston Rd, London NW1 2DB, UK}

\author{Ulrike Kuchner}
\affil{Centre for Astronomy and Particle Physics, University of Nottingham,
Nottingham, UK}

\author{Clár-Bríd Tohill}
\affil{Centre for Astronomy and Particle Physics, University of Nottingham, Nottingham, UK}

\begin{abstract}
Being able to distinguish between galaxies that have recently undergone major merger events, or are experiencing intense star formation, is crucial for making progress in our understanding of the formation and evolution of galaxies. As such, we have developed a machine learning framework based on a convolutional neural network (CNN) to separate star forming galaxies from post-mergers using a dataset of 160,000 simulated images from IllustrisTNG100 that resemble observed deep imaging of galaxies with Hubble. We improve upon previous methods of machine learning with imaging by developing a new approach to deal with the complexities of contamination from neighbouring sources in crowded fields and define a quality control limit based on overlapping sources and background flux. Our pipeline successfully separates post-mergers from star forming galaxies in IllustrisTNG 80\% of the time, which is an improvement by at least 25\% in comparison to a classification using the asymmetry (A) of the galaxy. Compared with measured S\'ersic profiles, we show that star forming galaxies in the CANDELS fields are predominantly disc-dominated systems while post-mergers show distributions of transitioning discs to bulge-dominated galaxies. With these new measurements, we trace the rate of post-mergers among asymmetric galaxies in the universe finding an increase from 20\% at z=0.5 to 50\% at z=2. Additionally, we do not find strong evidence that the  scattering above the Star Forming Main Sequence (SFMS) can be attributed to major post-mergers. Finally, we use our new approach to update our previous measurements of galaxy merger rates $\mathcal{R} = 0.022 \pm 0.006 \times (1+z)^{2.71\pm0.31}$

\end{abstract}


\section{Introduction} \label{sec:intro}
 
The first deep Hubble Space Telescope (HST) images of the distant universe revealed that many distant and faint galaxies are in fact irregular/peculiar in appearance \citep[e.g.,][]{  1996AJ....112.1335W}. Because the first cameras on HST, WFPC1/WFPC2 were sensitive in optical wavelengths only, probing distant galaxies was limited to their rest-frame ultra-violet light, due to the effects of redshift. It was unclear whether the peculiar appearances were the result of observational limitations or real. The question thus remained whether the observed irregularities were in fact just the star forming areas of these galaxies, while the older stars remained below detection. When the NICMOS camera was launched in 1998 on HST it became clear that the morphologies of distant galaxies were peculiar in their rest-frame optical wavelengths as well, implying that the bulk stellar mass in these galaxies was indeed out of equilibrium \citep[e.g.,][]{2000ApJ...531..624D, 2005ApJ...631..101P, 2005ApJ...620..564C, Mortlock2013, Whitney2021}. The common census was that distant galaxies are indeed intrinsically peculiar. However, it remained unclear why and how this finding relates to the various possible modes that could be responsible for producing these irregularities in galaxy structures at high redshifts. The peculiar appearance is likely linked to the formation process of the galaxies, but details of the origin of the observed irregular structures have proven difficult to fully understand. 

Since then, it has become clear that, overall, galaxies gradually transition from peculiar galaxies at higher redshifts to ellipticals and disc systems at lower redshifts  \citep[e.g.,][]{Conselice2003, Lotz2004, 2013MNRAS.433.1185M, 2015ApJS..221....8H}. This conclusion was made possible with the advent of the WFC3 camera on HST which allowed astronomers to trace the morphological evolution of galaxies over large areas of the sky. Galaxies are therefore undergoing a transformation, and their irregular origins reveal clues about the processes which drive galaxy formation.  
One popular and well explored hypothesis is that these systems are in fact undergoing hierarchical mergers to form larger systems.  The basic idea is that two galaxies in the early universe smash together to form a larger galaxy, a process which is predicted to be a critical element in the cosmological context of galaxy formation within a Cold Dark Matter (CDM) universe, with well defined predictions of this process \citep[e.g.,][]{2009ApJ...697.1971J, 2009MNRAS.396.2345B, Mundy2017}.  

To make progress in understanding the evolution of galaxies, it is crucial to identify merging galaxies correctly. In order to separate galaxies into mergers and non-mergers -- initially focusing on the nearby universe -- quantitative morphology tools were developed that use parameters such as the asymmetry index ($A$) \citep[e.g.,][]{2000ApJ...529..886C, Conselice2003a}. Merging galaxies are often identified through a combination of these morphology measurements such as the CAS parameters, which links measurements of the Concentration, Asymmetry, and Smoothness \citep[e.g.,][]{Conselice2003, Lotz2004}. However, mergers are not uniquely identifiable in this parameter space and some do not fall into the selection criteria at all \citep[e.g.,][]{Conselice2006, Lotz2008}. Therefore, care has to be taken to calibrate their usage.

Today, the merger rate can be accurately measured to high redshifts ($z \sim 3$) using galaxy structure \citep[e.g.,][]{Conselice2003, Conselice2008, Man2016, Mantha2018, FERREIRA2020, Whitney2021}. Using e.g., CAS parameters, the measurements show that the merger rate increases at higher redshifts up to $z \sim 3$, such that f$_{\rm merger} \sim (1+z)^{2-3}$ \citep[e.g.,][]{Conselice2014}, an evolution which scales similarly to the density of the universe, which evolves as $\sim (1+z)^{3}$.  This implies that with identifications of mergers at both high and low redshifts, we are able to trace the galaxy merger history and investigate the role of mergers within the formation of galaxies over time \citep[e.g.][]{Conselice2006, Mundy2017}.

In addition to high merger rates, distant galaxies have much higher star formation rates than today, peaking at $z \sim 2$ \citep[e.g.,][]{Madau2014}. We further know that galaxy structure is highly dependent on the star formation rate in the sense that intensely star forming galaxies generally appear more clumpy and irregular than quiescent galaxies at all redshifts \citep[e.g.,][]{Windhorst2002, MagerTaylor2018, Gao2018, Sazonova2021}. In fact, these two different types of galaxies -- mergers and non interacting intensely star forming galaxies -- can look very similar by eye, which complicates visual classifications. Even kinematically it can be challenging to distinguish mergers from rotating galaxies with high dispersions  \citep[e.g.,][]{2019ApJ...874...59S, Bottrell2021}. In addition, classifications and selections of galaxies after a merger event (post-mergers) are highly contaminated by misclassified isolated galaxies with high specific star formation rates (sSFR). This is because their star forming regions and dusty inter stellar medium (ISM) can generate asymmetric features reminiscent of (post-) merger features.
It is therefore currently unknown if and how we can correctly distinguish whether a galaxy is undergoing intense star formation, or some type of merger using galaxy structures and morphologies.  

\medskip

One way to approach this question is through novel techniques using machine learning. Recently, tremendous progress has been made in applying supervised deep learning methods to investigate galaxy morphology \citep[e.g.][]{Huertas-Company2018, Reiman2019, Huertas-Company2019, Cheng2018, Martin2019, Walmsley2020, Walmsley2021}. These end-to-end techniques are also very promising for investigating galaxy mergers specifically \citep{Ackermann2018, Pearson2019, Pearson2019a, Bottrell2019, Wang2020, FERREIRA2020, Bickley2021}. Additionally, one can also leverage information not only from visual classifications and observations, but also by forward modeling cosmological simulations to the observational domain \citep{deepMergeI, deepMergeII}.

We have recently started a machine learning exercise to determine the merger history of galaxies using cosmological simulation runs from IllustrisTNG \citep{Vogelsberger2014, Pillepich2018, Nelson2019}. In \cite{FERREIRA2020} we were able to separate mergers from other types of galaxies in IllustrisTNG to a success rate of 90\% up to $z \sim 3$.  The present paper is a followup to our first paper, in which we now investigate whether it is possible to distinguish merging galaxies from intensely star forming galaxies (above the SFR-stellar mass main sequence). These galaxies have the lowest success rates in classifications from \cite{FERREIRA2020}. Our task in this paper is to correctly distinguish mergers from star forming galaxies by only using their morphology and structure. 

This paper is organized as follows: in \S \ref{sec:selection} we describe the data sets we constructed for this task, from IllustrisTNG (simulations) and CANDELS (observations). A description of the methods we used to train a Deep Learning model and how we measure the structure of the galaxies in our samples is given in \S \ref{sec:methods}. We present our results in \S \ref{sec:results} while a discussion on the implications is laid out in \S \ref{sec:discussion}. Finally, we summarize and conclude our findings in \S \ref{sec:summary}.


\section{Data}\label{sec:selection}

To test our new Deep Learning approach, we use simulated galaxies from cosmological simulations post-processed with the SKIRT \citep{SKIRT8, Camps2020SKIRTGrains} dusty radiative transfer code. The simulations are based on IllustrisTNG (Sec. \S \ref{sec:illustris_data}) which are used for the construction of the training sample for a Convolutional Neural Network (CNN) that is subsequently applied to observed galaxies from the CANDELS fields (Sec. \S \ref{sec:CANDELSFIELDS}). Our sample definitions for post-mergers and star forming galaxies are given in Sec. \S \ref{sec:sample_definitions}.
We discuss the pipeline used to generate CANDELIZED mock images from IllustrisTNG in Sec. \S \ref{sec:pipeline_imaging_data}.

\subsection{IllustrisTNG}
\label{sec:illustris_data}

IllustrisTNG is a suite of cosmological, gravo-magneto-hydrodynamical simulation runs with a diverse set of particle resolutions. From highest to lowest resolution, simulations were realized in three comoving simulation boxes of $50, 100, 300 \rm \ Mpc \ h^{-1}$ length size, aptly named TNG50, TNG100 and TNG300 \citep{PillepichI, Naiman2018, Nelson2018a, Springel2018, Marinacci2018, Nelson2019, Pillepich2019}.

Our analysis makes use of the TNG100-1 simulation, which has proven to be a good compromise between resolution and volume\footnote{We tested our selections on TNG50-1 but resulting samples are too small for Deep Learning training.}. TNG100 has already been used extensively in studies that analyze galaxy morphologies and structures, including the comparison between simulations and observations \citep{HuertasCompany2019, Blumenthal2020}, and the use for deep learning \citep{Wang2020, Bickley2021, Bottrell2022}. Specifically, \cite{Zanisi2021} show that TNG100 galaxies reproduce observed objects well, especially disc-dominated sources. While there are some deviations in the small-scale structure of highly concentrated spheroidal systems, this is a minor issue in our analysis since they only make up a small fraction of our sample. In addition, our galaxies are resolution limited at the current redshift of interest, meaning that tiny details of structure are not relevant in this analysis.

To counterbalance any limitations from resolution we limit our analysis to galaxies with $M_* > 10^{9.5} M_\odot$.  Above this limit, and at our explored redshift range $z > 0.5$, galaxies are represented by thousands of stellar particles. This enable sampling the simulated galaxies into resolutions comparable to that of the observed CANDELS data. Specifically, the gravitational softening length of the simulation, $\epsilon$, is not a limitation when compared to the HST ACS and WFC3 cameras resolution.

This approach is a noticeable refinement to our previous treatment in \cite{FERREIRA2020}, where the research question did not demand the resolution of fine morphological features like clumpy regions and tidal features which are required for the present analysis. 

To select appropriate galaxies from TNG100-1, we isolate galaxies with  $M_* > 10^{9.5} M_\odot$, in the redshift range $0.5 < z < 3$. To limit contamination in our sample, we use a minimum dark matter to total mass ratio of

\begin{equation}
    \frac{M_{\rm DM}}{M_{\rm total}} > 0.1,
\end{equation}
as a way to avoid subhalos created as a result of disk fragmentation. This means that at least 10\% of the subhalo's mass needs to be in the form of dark matter. 
We acknowledge that this could also inadvertently remove galaxies that had their dark matter stripped, however this number is small and does not impact the final sample. This criteria removes $\approx 2\%$ of galaxies from the overall pool of available sources (subhalos) in TNG100.

We also remove objects that are smaller than the ACS PSF size from the selection. To identify these objects, we first convert the half mass radius $R_{1/2}$ provided  in the simulation group catalogs in kpc, to a pixel scale based on the cosmological model adopted by IllustrisTNG and the ACS pixel scale

\begin{equation}
    R_{1/2 Mass, pix} = \frac{a(z)}{h} R_{1/2 Mass, kpc}, 
\end{equation}
where $a(z)$ is the angular size at $z$ and $h$ the Hubble constant $/ \ 100$. Any galaxy with $R_{1/2 Mass, pix} < 3 \ \rm pix$ was then filtered out from our selections. This step removes $\approx 3\%$ of galaxies from the total pool of available sources.

\subsection{Sample Definitions}\label{sec:sample_definitions}

Our goal is to separate star forming galaxies from post-mergers at intermediate to high redshifts based on their morphology.
We define post-mergers as galaxies with at least one major merger event with a mass ratio 

\begin{equation}
    \mu = \frac{M_2}{M_1},\quad \mu \ge 0.25,
\end{equation}

where $M_1$, $M_2$ are the stellar masses of the galaxy pair involved in the merging event, ranked by their stellar mass respectively, with $M_1 > M_2$. Galaxies are considered post-mergers if they have coalesced into a single galaxy in the past $500$ Myrs, where a single galaxy is represented by a subhalo in the simulation as identified by friends-of-friends algorithms \citep{Rodriguez-Gomez2015}. This selection window timescale is motivated by the observability timescales of disrupted structures caused by mergers identified by structure measurements in IllustrisTNG \citep{Whitney2021}, and are higher than what was previously used in \cite{FERREIRA2020}. We allow post-mergers to have low sSFRs. Their asymmetric features likely arise from the merging process rather than from star forming clumps. In contrast, non-interacting star forming galaxies are defined here as galaxies that have sSFRs above the following threshold,

\begin{equation}
    sSFR > 10^{-9.5} \rm \ yr^{-1}
\end{equation} 

\noindent{and} are not interacting with other galaxies. To isolate non-interacting cases, we exclude any galaxy from the simulation that had major or minor merger events ($\mu > 0.1$) around $\pm 1$ Gyr of its current redshift. 
Minor mergers are excluded completely from both definitions, and any conclusions presented in this paper should be considered with this in mind. Importantly, this selection is not intended to limit the non-interacting cases to extreme starbursting episodes alone, but to select non-interacting galaxies with sufficiently high sSFR to produce clumpy and asymmetric features that could be mistaken for merging signatures.

In summary, this selection results in a sample of $\sim 6,000$ post-mergers and $\sim 110,000$ non-interacting star forming galaxies. While this may be a realistic representation of actual fractions (only $\sim 5\%$ of the sample are post-mergers), training the network requires a balanced dataset. We thus use the post-merger sample as the baseline and separate it in bins of redshift, stellar mass, and size, randomly sampling the same number of non-interacting galaxies within each bin. We remove bins without adequate matched numbers of star forming galaxies. This becomes noticeable in the higher mass bins where post-mergers dominate and very few star forming galaxies are present. 

\begin{figure*}
\label{fig:stats}
    \includegraphics[width=\textwidth]{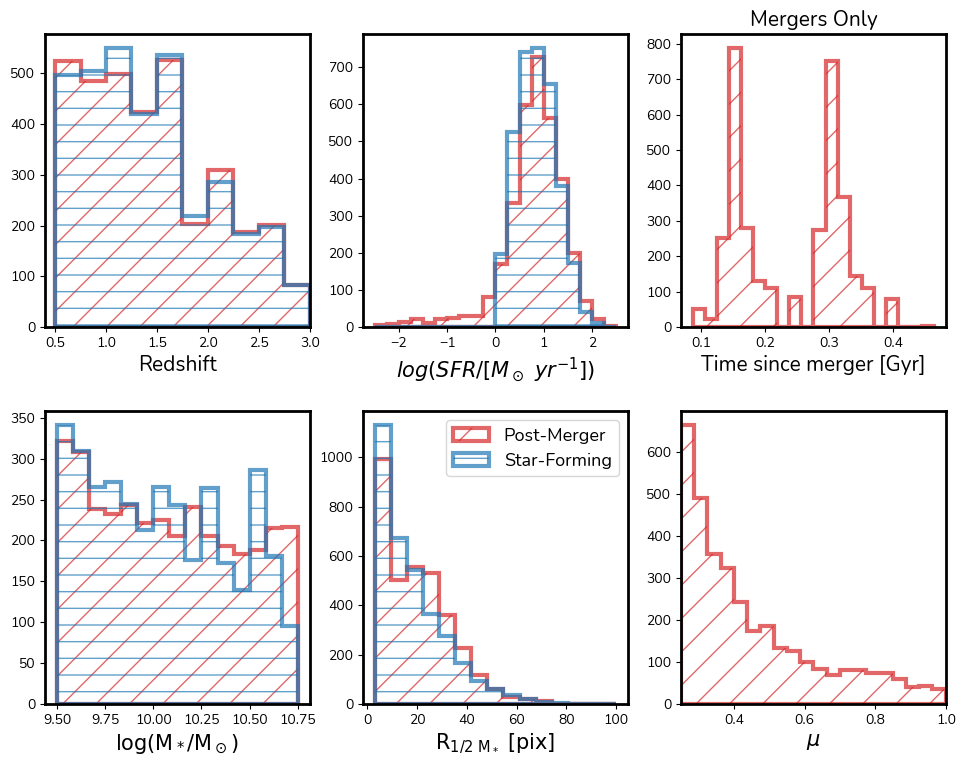}
    \caption{Physical properties of the 8,000 IllustrisTNG TNG100-1 selected simulated galaxies. For both types of galaxies we show distributions for redshifts (top left), star formation rates (top middle), stellar masses (bottom left) and stellar half-mass radius (bottom middle) in red for post-mergers, and blue for star forming galaxies. Distributions agree in general, with a small excess of stellar mass and size for the post-mergers. The time since the last major merging event and the mass ratio, $\mu$, -- properties unique to the post-mergers -- are shown in top right and bottom right, respectively.}
    \label{fig:sample_stats}
\end{figure*}

After matching the samples, we count $\sim 4,000$ galaxies in each class as our final sample. A summary of this sample separated by class and redshifts is available in Table \ref{tab:sample_sum}. The distribution of redshifts, star forming rates, stellar masses, and stellar half-mass radius are shown in Figure \ref{fig:stats} for post-mergers in red and star forming galaxies in blue. Both classes have very similar physical properties, with a small excess of large, passive and massive post-mergers in comparison to the star forming galaxies. Additionally, the top right and bottom right panels of Figure \ref{fig:sample_stats} show the time since the last major merger event, $\tau$, and the mass ratio, respectively, for post-mergers. The nature of the distribution for $\tau$ arises from the average time between snapshots in the simulation of around $\sim 0.15 \ \rm Gyr$. This timescale represents 1 to 3 snapshots after the coalescence of stellar masses.

\begin{table}[]
    \centering
    \begin{tabular}{cccc}
        \hline
             \textbf{Redshift} &   \textbf{Post-Mergers} & \textbf{Star Forming} & \textbf{Total}\\
             
             \hline
             $ 0.5 \le z < 1.0$ & 1214  & 1167 & 2381 \\ 
             $ 1.0 \le z < 1.5$ & 1082  & 1140 & 2222  \\ 
             $ 1.5 \le z < 2.0$ & 847  & 881 & 1728 \\ 
             $ 2.0 \le z < 2.5$ & 589 & 556 & 1145  \\ 
             $ 2.5 \le z < 3.0$ & 333  & 321 & 645  \\
    \end{tabular}
    \caption{Summary of the initial IllustrisTNG sample. The numbers in this table present the sample before each galaxy was post-processed with SKIRT and CANDELIZER (see text for detail), during which each image was augmented by 20 for 4 orientations and 5 different fields.}
    \label{tab:sample_sum}
\end{table}{}

\

\subsection{CANDELS Fields}\label{sec:CANDELSFIELDS}

One of the main goals of this work is to predict star forming and post-mergers galaxies in the observed CANDELS imaging data \citep{Grogin2011, Koekemoer2011}, which comprises high-quality HST observations from COSMOS, UDS, EGS, GOODS-South, GOODS-North \citep{Grogin2011,Koekemoer2011}. CANDELS data has been used extensively for galaxy merger studies, with estimated merger rates up to $z \sim 6$ \citep[e.g.,][]{Mantha2018, Duncan2019, Whitney2021}. Importantly, CANDELS also provides visually classified morphologies \citep{Kartaltepe2015}, as well as photometric redshifts, star formation rates, and stellar mass estimates \citep{Conselice2007, Duncan2014, Duncan2018a, Duncan2018b, Duncan2019} from SED fitting, essential for creating matched samples. Stellar masses are estimated within a $0.2 \rm \ dex$ systematic uncertainty, and the outlier fraction of photometric redshifts is smaller than $5\%$. We do not account for these uncertainties directly in our methods, but we refer the reader to the detailed discussion on the uncertainties associated with these physical measurements in the aforementioned references. Furthermore, the depth of the wide field data is comparable to the detection limit of galaxies in the IllustrisTNG simulations in the stellar mass range studied here. 

To select CANDELS galaxies we first remove all problematic objects according to their quality flags as recorded in the photometric catalogue and in the \cite{Kartaltepe2015} catalogues to avoid edges, artifacts and stars. Following \cite{Huertas-Company2016} and \cite{Kartaltepe2015}, we then select galaxies with H-band magnitudes $H < 24.5 \rm \ mag$. Because this cut can bias our sample against extended sources, we also include a signal-to-noise ($SNR$) lower limit of $\rm SNR > 50$ to exclude any compact source with only a few bright pixels. This magnitude cut removes 1074 sources, while the SNR cut further removes 430 sources. Then we proceed with the same cuts we used to select IllustrisTNG galaxies, using $0.5 < z < 3$ and $M_* > 10^{9.5} M_\odot$. We apply a final cut using the asymmetry ($A>0.1$, Sec. \S \ref{sec:morphology}) to remove regular unambiguous galaxies with no apparent disturbed or asymmetric features. This ultimately results in a sample of 23,494 galaxies from all the CANDELS fields combined.

Finally, we produce cutouts for $I_{814}$, $J_{125}$ and $H_{160W}$ bands centering on each selected CANDELS galaxy, each with a field of view of $50 \ \rm kpc \times 50 \rm \ kpc$, using photometric redshifts from \cite{Duncan2019}, preserving relative sizes between galaxies. Importantly, this selection does not rely on size measurements that could easily be spurious in interacting or merging galaxies. We do not find any bias in our classifications that could be attributed to small changes of the field of view caused by the photometric redshift uncertainties.

\subsection{Pipeline to produce CANDELIZED Mocks}\label{sec:pipeline_imaging_data}
\begin{figure*}
    \centering
    \includegraphics[width=0.9\textwidth]{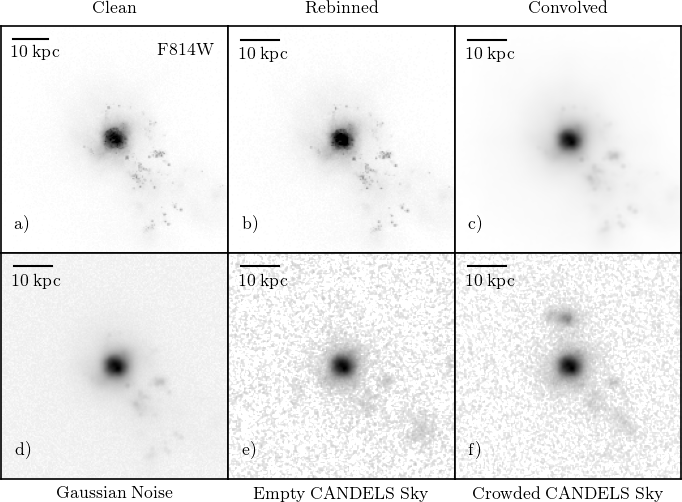}
    \caption{Example of the processing steps of our mock pipeline. \textbf{a)} Noiseless F814W broadband image generated from the simulated galaxy datacube with 0.03 $''/pix$ pixel scale. \textbf{b)} The same image after rebinning from z=0.5 to z=0.6. \textbf{c)} Image convolved by the HST F814W PSF. \textbf{d)} Image with Gaussian noise added. \textbf{e)} Image added on top of a random patch of the sky within a CANDELS field with no neighbouring sources. \textbf{f)} Image added randomly to a patch of sky with other sources in the field of view. As this patch of the sky is randomly selected, all final images have varying levels of contamination from nearby sources. We quantify this by the total flux in the sky patch before adding the simulated source to it.}
    \label{fig:example_mock_pipeline}
\end{figure*}

In order to guarantee realistic representations of CANDELS galaxies in the simulated sample, we must include instrumental and cosmological effects to the images of the IllustrisTNG galaxies. An overview of the steps are shown in Fig. \ref{fig:example_mock_pipeline} and are detailed in this section. 
IllustrisTNG data holds information on the stellar, gas and dark matter particles for each source. Each particle represents a large physical region that can be described by rich stellar populations, that vary depending on age, mass and metallicity. The resampling of the star forming regions is particularly important to avoid problems with the coarse representations \citep{CAMPS2016, Trayford2017OpticalSKIRT}.

To create mock broadband images, we thus process each stellar particle with a population synthesis model 
following the recipes from \cite{Trayford2017OpticalSKIRT} \& \cite{ Vogelsberger2020High-redshiftFunctions}. This entails post-processing the simulation data with the Monte Carlo dusty radiative transfer code SKIRT \citep{SKIRT8, Camps2020SKIRTGrains}. 

Each stellar particle in the simulation is considered as a Single Stellar Population (SSP) with GALAXEV \citep{BruzalCharlot} or MAPPINGSIII  \citep{MAPPINGS} SEDs based on its stellar mass, absolute metallicity, and age. We choose to adopt these particular templates because, firstly, they are implemented in SKIRT and, secondly, they had been tested previously in similar pipelines to generate mock observations from cosmological simulations \citep{Trayford2017OpticalSKIRT, Rodriguez-Gomez2018}. Finally, \cite{BruzalCharlot} are also the templates used to derive stellar masses and star formation rates for all the CANDELS fields in \cite{Duncan2019} which are used in this study.

To account for the fact that each stellar particle represents an extended area (rather than treating them as a point source), we model the particles with a smoothing length of a truncated Gaussian emissivity profile equal to the distance to its 64th neighbor particle \citep{Trayford2017OpticalSKIRT}. We then define a grid of wavelengths covering all spectral features we want to probe within the HST filter response functions, similar to the grid used in \cite{Trayford2017OpticalSKIRT}. For each wavelength bin of this grid, we launch $10^6$ photon packets, assuming isotropic emission until they reach the virtual detector.

\begin{figure*}
    \includegraphics[width=\textwidth]{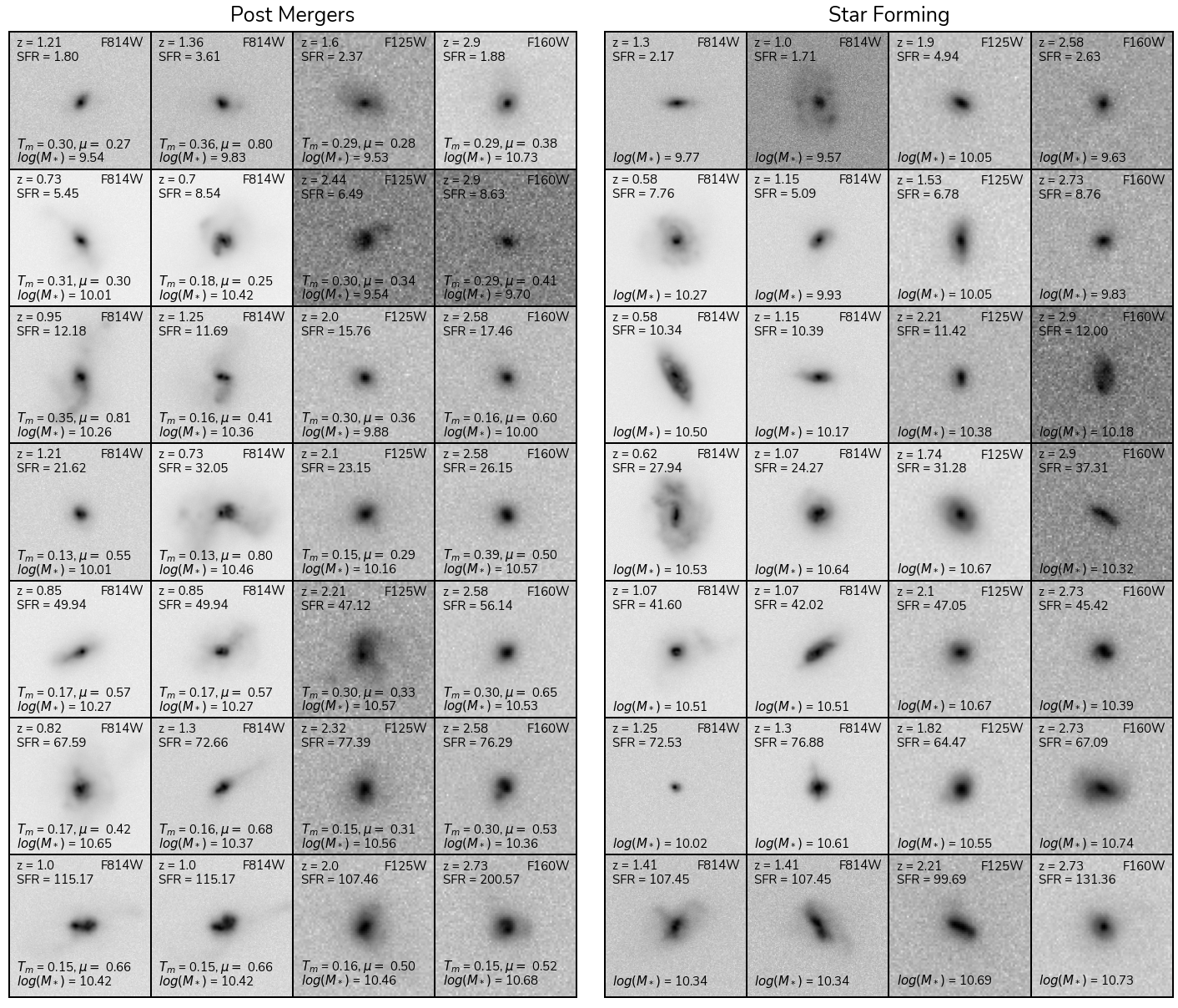}
    \caption{A random selection of IllustrisTNG simulated galaxies in our test sample; post-mergers (left) and star forming galaxies (right), with their redshifts, SFRs, and stellar masses printed in each stamp. Images are ordered from left to right in redshift, and top to bottom in SFR. For post-mergers we also display the time since merger, $T_m$, and the mass ratio $\mu$. All stamps use a square-root normalization.}
    \label{fig:mosaic_illustris}
\end{figure*}

\begin{figure*}
    \centering
    \includegraphics[width=\textwidth]{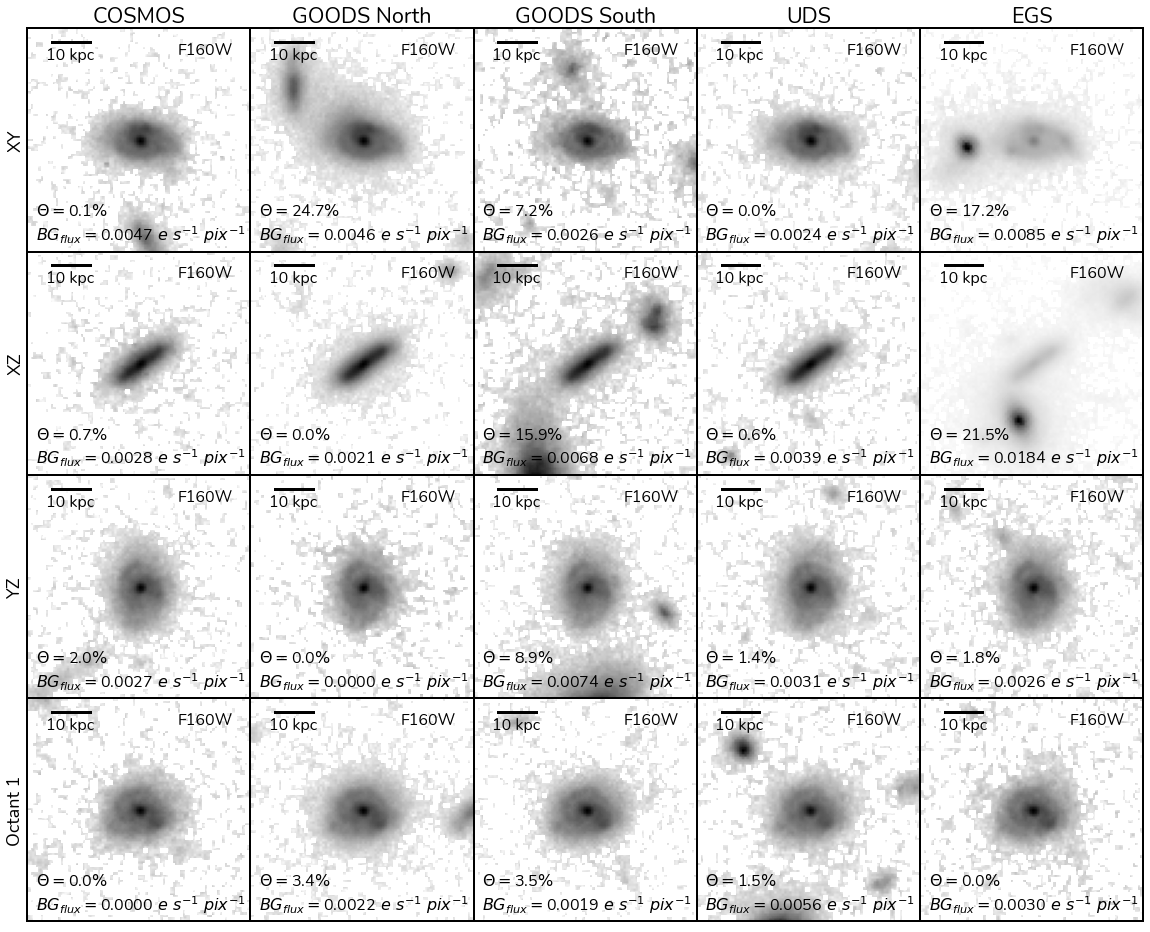}
    \caption{Demonstration of the augmentation pipeline for one random galaxy from TNG100-1 (ID=192802, z=0.55, at different orientations). We increase our sample by augmenting the dataset, reproducing it in four orientations (rows) in each of the CANDELS fields (columns). The simulated galaxy is placed in a random patch of the sky in the CANDELS fields and thus can have other sources in the final cutout. The amount of contamination from neighbouring sources varies widely due to the random sampling of the background described in \S \ref{sec:pipeline_imaging_data}. This contamination is quantified by the overlapping percentage, $\Theta$, and the average flux of the background patch, $BG_{flux}$}
    \label{fig:augmentation_mosaic}
\end{figure*}
\bigskip

This process produces IFU datacubes over the SKIRT wavelength grid which we then reduce to broadband images with the same properties as the CANDELS HST images.
SKIRT's reference frame used to generate the datacubes is located at a distance of 10 Mpc (initial redshift $z_0$) of the sources. We must therefore shift the IFU data to each target's redshift, $z_{t}$, by $(1+z_{t})$ while dimming its flux by 
\begin{equation}
    \frac{f_t}{f_{o}} = (1+z_t)^{-1} \left (\frac{D_{Lo}}{D_{Lt}} \right )^2,
\end{equation}
due to cosmological dimming \citep[eq. 15]{Hogg1999}. Next, we convolve the IFU data with the broadband filters response functions for $I_{814}$, $J_{125}$ and $H_{160W}$. The results are clean, noiseless images from the simulation galaxies at $30 {\rm \ mas / pix}$ (matching the ACS pixel scale) before adding any PSF effects (Fig. \ref{fig:example_mock_pipeline}, images a) to c). 
We rebin $J$ and $H$ bands from $30 \ {\rm mas / pix}$ to the WFC3 images pixel scales of $60 \ {\rm mas / pix}$. Examples for stamps where the background was added can be seen in Fig. \ref{fig:example_mock_pipeline}, images e) and f).

Figure \ref{fig:mosaic_illustris} gives randomly selected examples of galaxies in our sample before any contamination from the CANDELS sky is included, separated by their class.

\bigskip
The data-driven paradigm of Deep Learning methods imposes high requirements on the amount of data necessary to train a model that is capable of generalizing the training data well. In practice this means that for the majority of models, a successful approach requires tens, hundreds or even millions of examples. We are far away from these numbers in  cosmological simulations. Our initial selection results in a balanced set of $\sim 4000$ examples of each class (Sec. \ref{sec:illustris_data}). Fortunately, in the case of galaxy images there are ways to increase the initial dataset by exploiting aspects of the final image that do not depend directly on the simulated galaxy. In our case, we apply data augmentation to our dataset in three ways outlined below. An example of this approach is shown in Fig. \ref{fig:augmentation_mosaic}, following the same galaxy in each possible combination of orientation/field. 

First, since IllustrisTNG provides the 3D distribution of all particles associated with a galaxy, we generate each galaxy with different line of sight projections, treating each new representation as a new galaxy. We select four different projections, three aligned with the axis of the simulation, XY, XZ and YZ, respectively, and a fourth line of sight aligned with one octant of the simulation cube.

Secondly, each CANDELS field has unique observational properties (e.g. different noise levels, depth). We exploit this aspect and reproduce each of the different orientations from the previous step on top of a random patch of sky of each CANDELS field, taking care to use appropriate noise levels for the simulated galaxy. To find empty patches of sky, we randomly sample the RA and DEC within each field, and make a large cutout of the area that is 4 times larger than the final size of the cutout. Using positions given in the CANDELS catalogs, we then identify all sources within this cutout and reselect a new RA and DEC within the cutout that does not centrally overlap with another source. We allow some degree of overlapping source, but require a unique central position. We do this interactively until a patch of sky that matches all above criteria is found. This, combined with all the orientations, augments our data set 20 times. In addition, this also helps the network to generalize the impact of contamination from neighbouring sources, as the same galaxy in one field might be isolated in its cutout, but in a denser environment in another.

Finally, we apply random flips, rotations and small zoom-in/zoom-outs around the central source on the fly during training as a regularization technique. This does not increase the overall size of the sample, but at each training epoch the network sees different realizations of the same sample. 

Overall, our sample increases from $\sim 8,000$ examples to $\sim 160,000$. However, having multiples of similar galaxies in our dataset can result in overfitting. To reduce this risk, we do not allow different realizations of the same galaxy to fall in both the training sample and the test sample. This ensures that testing and validating are performed on unique datasets. 

\begin{figure}
    \centering
    \includegraphics[width=0.47\textwidth]{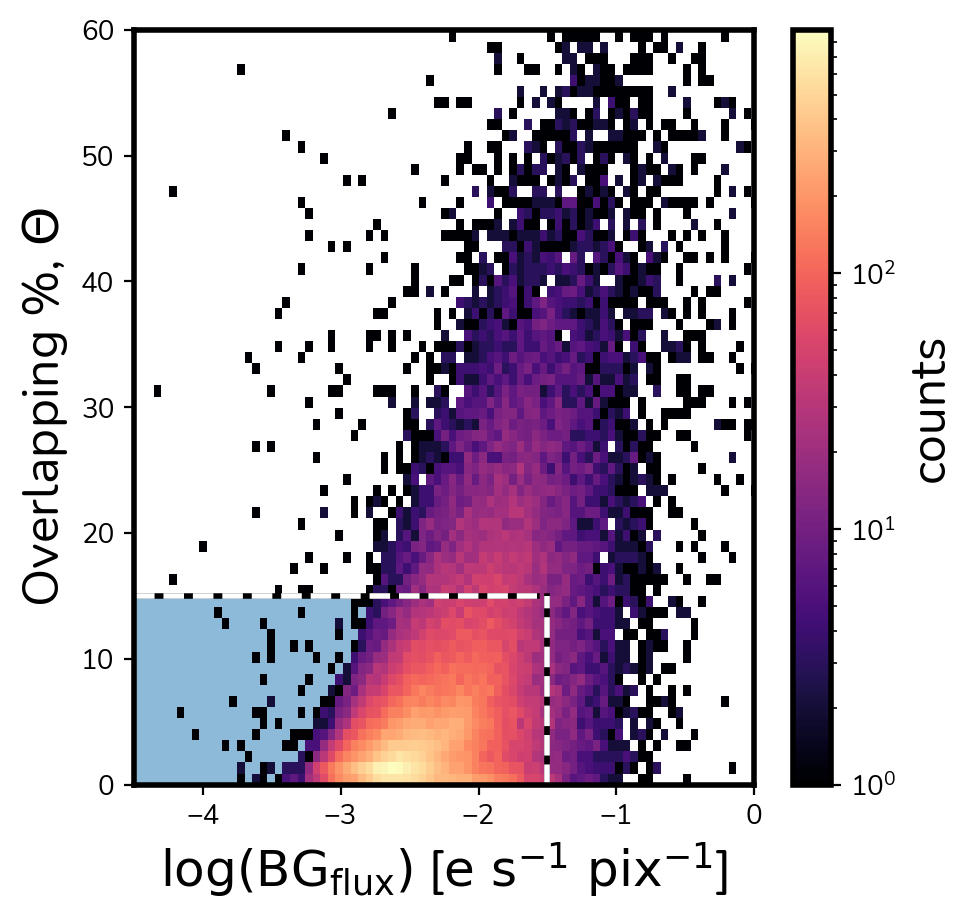}
    \caption{Contamination characterization for $162,000$ IllustrisTNG simulated images in our sample. We show the logarithm of the average flux per pixel of the background measured in each cutout, $\log(\rm BG_{flux})$ vs. the overlapping percentage, $\Theta$, which indicates how much the central galaxy segmentation map is covered by the segmentation map of the sources in the background. We define a conservative region of low contamination shown by the dashed line and blue area, which contains $90\%$ of the whole sample. Every point represents at least one image. }
    \label{fig:cover_FPP}
\end{figure}

\subsection{Contamination Quantification} 

\label{subsub:contquant}

Providing realistic levels of contamination and the inclusion of neighbouring sources are some of the most important requirements for a good generalization between samples of simulated galaxies and real observations \citep{Bottrell2019}. In an update to what was done in \cite{FERREIRA2020}, we included realistic contamination in our IllustrisTNG sample, as described in \S \ref{sec:pipeline_imaging_data}. By comparing clean galaxy realizations to their respective background-added images, we can thus test how our methods behave when faced with a variety of contamination levels, drawing direct conclusions for real world applications. We quantify the degree of contamination in each image using two measurements, which are also listed in Fig. \ref{fig:augmentation_mosaic}.

First, we define how much of the galaxy is covered by a background source. We call this the overlapping percentage, $\Theta$. For this we measure segmentation maps both for the central source and all background sources of each image stamp. $\Theta$ is the percentage of the segmentation map of the central galaxy that is covered by segmentation map(s) of background sources and ranges from 0\%, for no overlap, to 100\%, where the central galaxy is completely covered by another galaxy in the field.
 
Second, we estimate the average flux (per pixel) of all background sources, $\rm BG_{flux}$, by averaging the flux of the sources within the segmentation map over its area. $\rm BG_{flux}$ values are given in units of $e \ s^{-1} \ pix^{-1}$. This ranges from $\rm BG_{flux}\sim 0$, where there is no apparent or very faint source in the background, to values that are comparable to or even higher than the flux of the central source. Very high values may be due to bright neighbouring sources that outshine the central galaxy. Stars can also be identified by this method.

We use the overlapping percentage, $\Theta$, and the flux of background sources, $\rm BG_{flux}$, to define galaxy images with low contamination. Figure \ref{fig:cover_FPP} shows the parameter space formed by these two measurements for the entire sample of $\sim 160,000$ simulated and \textit{candelized} images. The blue box framed by the dashed line  defines a region of galaxies with low contamination,
\begin{equation}
    \Theta < 15\%, \quad \rm BG_{flux} < \times 10^{-1.5} \ e \ s^{-1} \ pix^{-1},
\end{equation}
which can be considered as a conservative choice. We find that $\sim 90\%$ of our sample is located in this region. We do not remove the remaining 10\% of the galaxies from our sample, because such highly contaminated cases will also be present in observations. We use these contamination estimates to understand how our methods are impacted by it.

These two properties form a simple and powerful way to characterize the contamination of our sample, as they control different contributions to contamination. Because these are challenging to measure directly in real CANDELS observations, we trained a deep learning model to predict the same values in real images. We describe this exercise in Appendix \ref{appendix:contamination}. By inference, any discussion based on contamination measurements in our simulation sample is also valid for the CANDELS observations.

\section{Methods}\label{sec:methods}

We use a deep learning framework with a Convolutional Neural Network (CNN) based on \cite{FERREIRA2020} but with significant updates related to the improved and more robust data pipeline that was discussed in \S \ref{sec:pipeline_imaging_data}. In this section we describe our Deep Learning analysis (\S \ref{sec:deeplearning}), where we also highlight the improvements to \cite{FERREIRA2020}. In Section \S \ref{sec:augmethod}, we discuss how to avoid overfitting due to the augmentation of the TNG sample, which was part of our sample pipeline. We further wish to compare the resulting classifications to ``traditional" classifications. We thus measure non-parametric morphology indices, structural parameters and S\'ersic profiles for both the TNG sample and the CANDELS sample with \textsc{Morfometryka} \citep{Ferrari2015, Ferreira2016, Lucatelli2018}, for which we provide a brief overview in Section \S \ref{sec:morphology}.

\subsection{Deep Learning Classifications}\label{sec:deeplearning}

We employ neural networks to forward model the simulations into the observational domain. The neural network takes galaxy images as input and outputs a probability associated with its classification, in this case whether it is a post-merger or a star forming galaxy. 

Neural networks are known for being able to approximate complex functions where no analytical approach is feasible, based on the universal approximation theorem \citep{ZhouApproximation}. Deep neural nets combine several layers of nodes (neurons) in a feed forward fashion, mapping inputs to outputs using non-linear activation functions.
As a data-driven method, the underlying rules are not explicitly programmed into the network but learned from pattern recognition on the relationship between inputs and outputs of data. These rules are found by minimizing a loss function between the true outputs and the predicted outputs. It is optimized by adjusting the weights and biases of the network so that the loss function reaches a minimum.

A Convolutional Neural Network (CNN) is an end-to-end method, where the most meaningful spatial features are also learned from the data itself through convolution operations. These features are then combined for a classification task, producing the desired outcome based on the input.

In this work, we use an improved version of the CNN architecture described in \cite{FERREIRA2020}. This consists of a feed-forward network with an input image size of 128x128 pixels, where the number of convolutional blocks, convolutional layers, fully connected layers, number of filters, and kernel sizes, are all defined by the following hyper-parameters:

\begin{itemize}
    \item \texttt{number\_conv\_blocks} define the number of convolutional blocks, each will probe features of different scales;
    \item \texttt{number\_conv\_per\_block} describe how many convolutions each block will have;
    \item \texttt{initial\_number\_filters} define the starting number of filters, that are then doubled after each convolutional block;
    \item \texttt{initial\_kernel\_size} is the initial size of the convolutional kernel, that is then reduced by 2 after each block, down to a minimum of 3;
    \item \texttt{n\_fc\_layers} and \texttt{size\_fc\_layers} define the number of hidden layers and their respective size, respectively; 
    \item \texttt{l2\_regularization} and \texttt{dropout} are the degrees for each regularization technique used, respectively. l2 regularization is applied to all convolutional layers, where dropout is applied only after the hidden layers;
\end{itemize}
\noindent The approach of variable depth and width for neural networks is similar to the family of networks described in \cite{efficientNet}. However, in our case the networks are smaller due to the smaller image size used.

We modify the methods from \cite{FERREIRA2020} to improve generalization of our models. First, instead of using two binary classification networks and combining their predictions to construct a multi-class classification, we now only use one network for the binary classification of post-mergers and star forming galaxies.

Second, we treat the learning rate differently. In \cite{FERREIRA2020}, we monitored the learning rate decays during training as a hyper-parameter. Here, we use Cosine Annealing, a type of learning rate scheduling \citep[e.g.,][for an explanation]{ML_CossineAnneling} combined with a regular Stochastic Gradient Descent (SGD) optimizer \citep{SGD_generalization}. This approach probes several different learning rate regimes during training and uses cyclic resets that serve as a way to avoid unstable local minima, improving generalization of the solutions. 

All hyper-parameters are determined by a Bayesian Optimization process \citep{gpyopt2016}, and the values for the best model used here are summarized in Table \ref{tab:architecture}. These values can be directly used in conjunction with our public \textsc{keras} implementation. 

\begin{table}[]
    \centering
    \begin{tabular}{|r|c|}
        \hline
         Hyperparameter & Best Model  \\
         \hline
          batch\_size  & 128 \\
          number\_conv\_blocks  & 3 \\
          number\_conv\_per\_block  & 2 \\
          initial\_number\_filters  & 32 \\
          initial\_kernel\_size   & 11  \\
         
          number\_fc\_layers & 2 \\
          size\_fc\_layers & 128 \\
          l2\_regularization  & 0.1 \\
          dropout & 0.5 \\
         \hline
         
    \end{tabular}
    \caption{The best hyper-parameters of our architecture found through Bayesian Optimization \citep{gpyopt2016}. These define the depth,  width, and number of trainable parameters of our architecture. This process is done using our set-aside validation samples. The same model is used for all the CANDELS datasets.}
    \label{tab:architecture}
\end{table}{}

\subsection{Augmentations and Overfitting Avoidance}\label{sec:augmethod}

To avoid overfitting pitfalls from using our CANDELS background augmentation pipeline (\S \ref{sec:pipeline_imaging_data}), we train a suite of models, one for each CANDELS field. Because we have included areas of all the CANDELS fields as background in our training set, the network could potentially memorize these and use them for predictions, impairing the results. To ensure this is not the case, each CANDELS field has two models -- one at low redshift, $0.5<z<1.5$, and one at high redshift, $1.5<z<3.0$ -- trained only with images augmented with regions of the other four fields. All datasets (training, validation and test) are restricted in this way, guaranteeing that any overfitting of the CANDELS background will have no impact in the final application of our models.

\begin{figure}
    \includegraphics[width=0.47\textwidth]{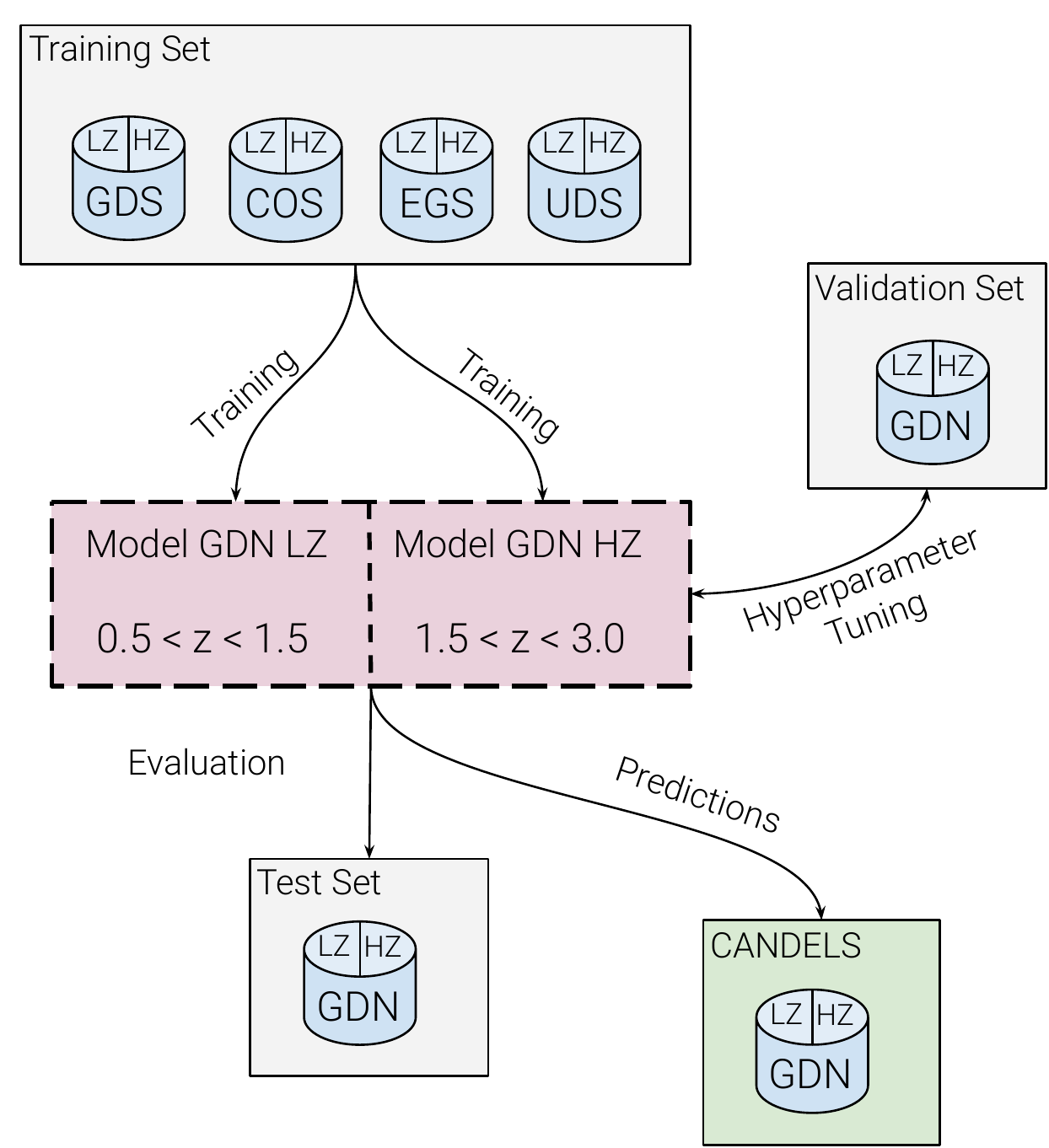}
    \caption{Schematics of the training pipeline leveraging multiple fields for augmentation. Each pair of models, at low redshift (LZ), and high redshift (HZ), is trained only with data that is augmented with the CANDELS fields that are not the target for the model. In this example we show a model designed for predictions on GOODS North (GDN), trained on data augmented with characteristics of all the remaining four fields (GDS, COS, EGS, UDS). This model is also tuned and evaluated in validation and test sets that have only of target CANDELS field augmentations, ensuring that no overfitting of neighbouring sources is part of the predictive process.}
    \label{fig:schematics}
\end{figure}

An example of this process is outlined in Figure \ref{fig:schematics}, for the models that will be used for predictions in the GOODS North (GDN) field. The training set contains galaxies augmented with the COSMOS (COS), GOODS South (GDS), Extended Groth Strip (EGS) and The UltraDeep Survey (UDS) fields while the validation and test sets only contain galaxies from GDN. 

This ensures that each model is tailored to one CANDELS field and that no source from that particular field is used during training, i.e., the network never sees any of its data. We further apply a regularization method that makes use of random rotations and image flips on the fly during the training time.

\subsection{Galaxy Structure and Morphology}\label{sec:morphology}

Non-parameteric structure measurements of galaxies are a traditional way to select galaxy mergers \citep{Conselice2003, Lotz2004, Lotz2008, Snyder2017}.
To measure structures for our sample, we fit S\'ersic profiles to all galaxies, using the software \textsc{Morfometryka} \citep{Ferrari2015, Ferreira2016, Lucatelli2018}. \textsc{Morfometryka} measures asymmetry ($A$) concentrations ($C$), the Gini coefficient ($G$), moment of light of the brightest pixels ($M_{20}$), normalized information entropy ($H$) and others. It also measures several structural parameters and fits 1D and 2D S\'ersic profiles. For our purpose, we are particularly interested in the asymmetry of the galaxies ($A$), as well as their smoothness  (S)  since, together, they define a common criterion for finding galaxy mergers: 
\begin{equation*}
    (A > 0.35), \quad (A > S). 
\end{equation*}
\noindent The asymmetry is defined as the pixelwise normalized difference between the original image and the same image rotated by 180 degrees,
\begin{equation*}
    A = \frac{\Sigma | I-I_{180}| }{\Sigma|I|} - A_{bg},
\end{equation*}
where $I$ is the image, $I_{180}$ is the rotated image and $A_{bg}$ is an asymmetry term associated with the background \citep[e.g.,][]{Conselice2014}. We measure $A_{bg}$ in each cell of a meshgrid overlayed onto the image, omitting the area occupied by the segmentation map of the central galaxy. We then use the median of these values as $A_{bg}$. This ensures a robust modeling of the impact of the background in the resulting asymmetry of the image \citep[e.g.,][]{CBMorpho}. 

Finally, as we are especially interested in investigating the nature of the peculiar/irregular cases, we follow the hybrid method proposed by \cite{Bickley2021}. We first filter out regular symmetric galaxies from the sample using the asymmetry ($A$). Instead of using the widely used cut for selecting mergers ($A>0.35$), we choose a conservative selection of galaxies with,
\begin{equation*}
    (A > 0.1).
\end{equation*}
This will remove  cases that are irrelevant for our research question. These are galaxies without any disturbances that would classify them as peculiar or irregular.

\begin{figure}
    \includegraphics[width=0.46\textwidth]{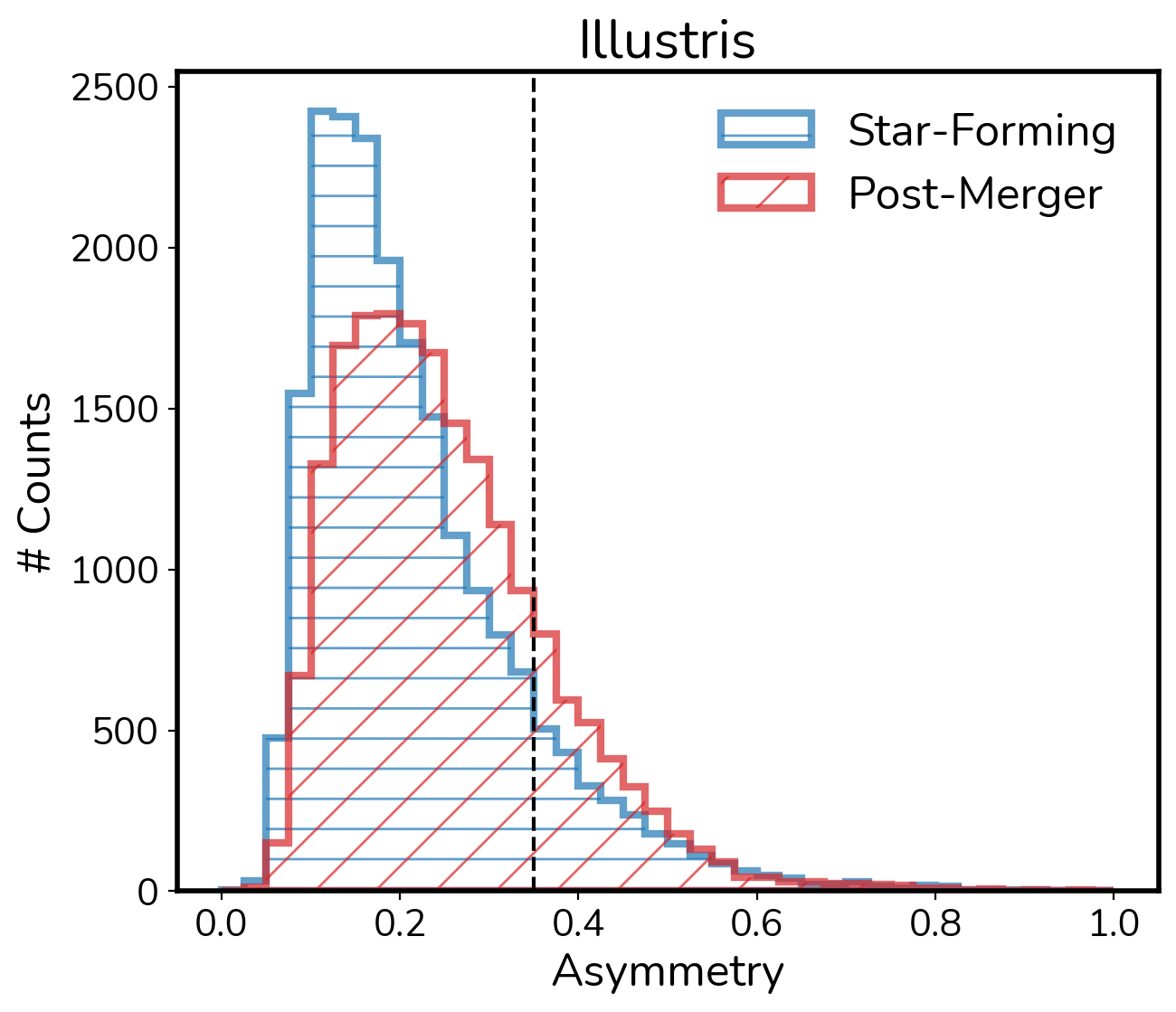}
    \caption{Distribution of asymmetries $A$ measured with \textsc{Morfometryka} for our TNG100-1 sample of galaxies. Star forming non-mergers and post-mergers are shown in blue and red, respectively. The dashed vertical line illustrate the typical threshold ($A>0.35$) used to classify galaxies as mergers.}
    \label{fig:A_illustris}
\end{figure}

In Fig. \ref{fig:A_illustris} we show the distribution of asymmetries $A$ measured with \textsc{Morfometryka} for star forming galaxies (in blue) and post-mergers (in red) for the simulated galaxies. The distributions largely overlap, though  asymmetries for post-mergers are generally slightly higher. The difference between both distributions is small enough that using solely the asymmetry ($A > 0.35$) will produce samples with low completeness and purity, and given that the fraction of merging galaxies is lower than regular star forming galaxies, it is likely that this approach produces very contaminated samples.

\section{Results}\label{sec:results}

Here we discuss what our trained models reveal, first from the test dataset of IllustrisTNG selected galaxies (\S \ref{subsec:pred_on_illustris}), and then applied to the CANDELS fields (\S \ref{sec:classifications_on_Candels}). 

\subsection{Predictions within IllustrisTNG}\label{subsec:pred_on_illustris}

\begin{figure*}
    \includegraphics[width=\textwidth]{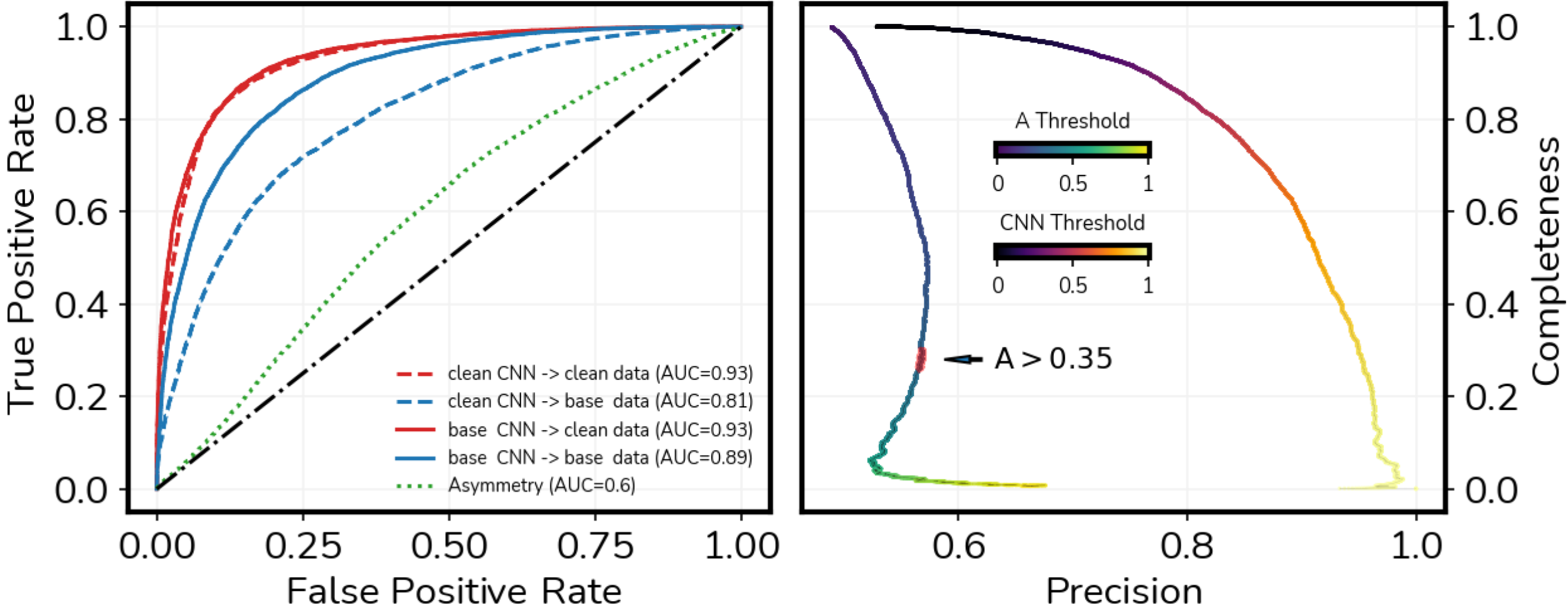}
    \caption{Performance metrics for our four trained models and comparison with the classical Asymmetry index $A$ for the simulated images. \textbf{Left:} ROC curves for both the network trained with the {\tt pristine mocks} dataset (dashed lines) and with the {\tt realistic mocks} dataset (solid lines) applied to both datasets, color coded in red ({\tt pristine}) and blue ({\tt realistic}). The green dotted line indicates the ROC curve for a classifier using only the asymmetry $A$. The area under each curve can be read in the label. \textbf{Right:} Precision-Completeness diagrams for the baseline network trained with the CANDELS matched mocks with asymmetry $A$, color-coded by classification threshold levels for CNN ({\tt inferno}) and asymmetry ({\tt viridis}). A small region in red is printed over the asymmetry curve to point out the region where the classification threshold is $A > 0.35$.}
    \label{fig:ROC}
\end{figure*}
We measure the performance of our trained models in our prepared test sets. This is done by training the network with two realizations of the test datasets, one with full HST-matched properties including a CANDELS background patch of the sky (\S \ref{sec:pipeline_imaging_data}, \S \ref{sec:augmethod}) (which we call \texttt{realistic mocks}) and one with clean mocks with no sky noise and contamination included (which we call \texttt{pristine mocks}). For simplicity, in cases where we only mention the \texttt{realistic mocks} without specifying which CANDELS fields it was augmented with, we consider the average of all 20 models described in \S \ref{sec:augmethod}.

To compare between models and realizations of these datasets, we use traditional performance metrics common for evaluating Machine Learning model performance. These consist of Receiver Operating Characteristic (ROC) curves and Precision-Completeness diagrams \citep{Powers2011}, as well as confusion matrices and their individual indices. 
Here, we are dealing with a single binary classification task, such that the probabilities of both classes respect the condition $\rm P(NMSF)+P(PM)=1$. Fig. \ref{fig:ROC} displays the overall performance for each network. 

The left panel shows four different realizations of the network for comparison purposes. The network is trained twice to generate two different types of models: one labeled {\tt base} that consists of a network trained with the {\tt realistic mocks}, and a second labeled {\tt clean}, which is trained with the {\tt pristine mocks}. Then, each model is applied to both datasets. We do this to measure the best case scenario within the simulations, in the absence of any contamination or impact from observational effects. Models trained with the {\tt realistic mocks} dataset are plotted as solid lines, while models trained with the {\tt pristine mocks} dataset are shown by dashed lines. Furthermore, the color conveys the dataset in which the model was applied to, red and blue for {\tt pristine mocks} and {\tt realistic mocks}, respectively. In addition to these, a single parameter classifier based on the asymmetry ($A$) is also evaluated and displayed as the green dotted line. The area under the curve (AUC) for each case can be found in the legend of the left panel.

The different realizations of our network ({\tt base} and {\tt clean}) cross-correlated with the {\tt realistic mocks} and {\tt pristine mocks} datasets confirms the importance of realistic observational modeling of the mocks (discussed in detail by \cite{Bottrell2019}). This is especially important when crossing domains from cosmological simulations to real observations. Figure \ref{fig:ROC}  shows that the {\tt base} network performs just as well as the {\tt clean} network when applied to the {\tt pristine mocks}, resulting in similar performance metrics, as can be seen by the overlapping red curves.  However, the {\tt base} network outperforms by $\sim 10\%$ the {\tt clean} network when applied to the {\tt realistic mocks} dataset, as displayed by the difference between the blue curves in Figure \ref{fig:ROC}. 
This demonstrates that correctly modeled observational features increase the generalization capabilities of the resulting models. A network that is only trained on pristine images will perform poorly in the real observations domain.

Importantly, all cases outperform the asymmetry by $20-30\%$. To some extent, this is expected because asymmetries of post-mergers are lower than asymmetries of galaxies that are just in the beginning of their merging event, including cases of closely interacting galaxies. Evidently, the asymmetry function is a much more general morphological descriptor while the network is very specialized for the particular task of dividing post-mergers from star forming galaxies. 

We compare the performance of asymmetry ($A$) and CNN predictions further and show completeness-purity diagrams in the right panel of Figure \ref{fig:ROC}. It displays outcomes for our ensemble of CNN models in {\tt inferno} colormap, and for the classic asymmetry parameter in {\tt viridis} corlormap. The  commonly used asymmetry value to classify galaxy mergers is generally higher than ($A > 0.35$), which is shown in the figure by the red patch over the curve. However, here we compare an asymmetry classifier with our neural network to exemplify how one can use the classification threshold of the network as a way to control the trade-off between precision and completeness. This is a useful feature when dealing with unbalanced datasets, like the case for galaxy mergers.  

The precision and completeness of the asymmetry behaves in unpredictable ways. First, the precision of the selection increases slowly, then it decreases again around ($A \sim 0.2$), and spikes above 0.6 precision for ($A > 0.8$), but with very low completeness. We do not seek to redefine its use, but merely contrast it with our deep learning approach, and show in broad terms when it might fail when dealing with ambiguous morphologies. 

\begin{figure*}
    \includegraphics[width=\textwidth]{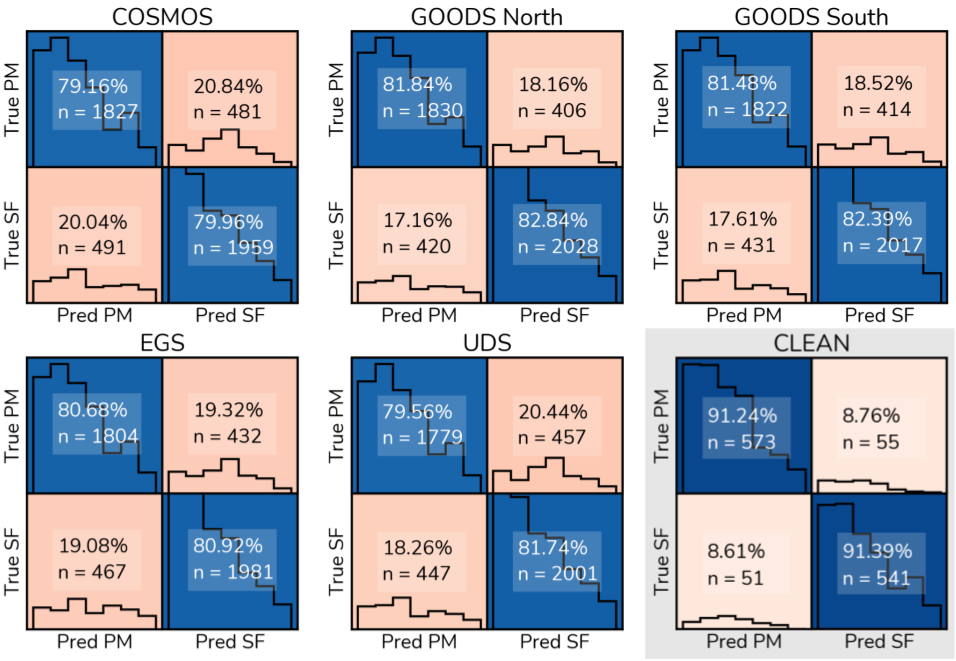}
    \caption{Confusion matrix for all the samples matched to CANDELS fields as well as the the {\tt pristine} sample (highlighted by gray shading in the bottom right). These confusion matrices were evaluated with the ensemble of models trained with the CANDELS matched mocks. We show True Negative (TN) and True Positives (TP) highlighted in blue while the False Negatives (FN) and False Positives (FP) are shown in pink. The colors are based on the rate percentage, which is also printed in each cell. All the CANDELS fields have TP and FN rates of around $\sim 80\%$. For the {\tt pristine} case performance can reach as high as $\sim90\%$, marking the intrinsic limit of our method based on the data available. The histograms show the redshift distribution for the galaxies in each category, which demonstrate that it is easier to recover correct classifications at lower redshifts.}
    \label{fig:vanillaCM}
\end{figure*}

Our network is able to correctly identify post-mergers and star forming galaxies from the IllustrisTNG simulation in $\sim80\%$ of the cases. Figure \ref{fig:vanillaCM} shows the confusion matrix for the {\tt realistic mocks} dataset identified within each individual  CANDELS field, as well as for the {\tt pristine mocks} sample, where accuracy reaches $\sim90\%$. All classifications are done with the model trained with the {\tt realistic mocks}. We show True Positives (TP) and True Negatives (TN) in blue, and False Positives (FP) and False Negatives in pink.   The {\tt CLEAN} case represents the best case scenario, where our current method and dataset achieves an even higher performance of $\sim91\%$ True Positives. A histogram of the redshift distribution for each cell helps to visualize any possible biases in redshift for the miss-classification cases. This demonstrates that the models are more likely to correctly classify low redshift galaxies, as they represent the majority of the samples.

\subsubsection{Impact of Redshift}

\begin{figure*}
    \includegraphics[width=\textwidth]{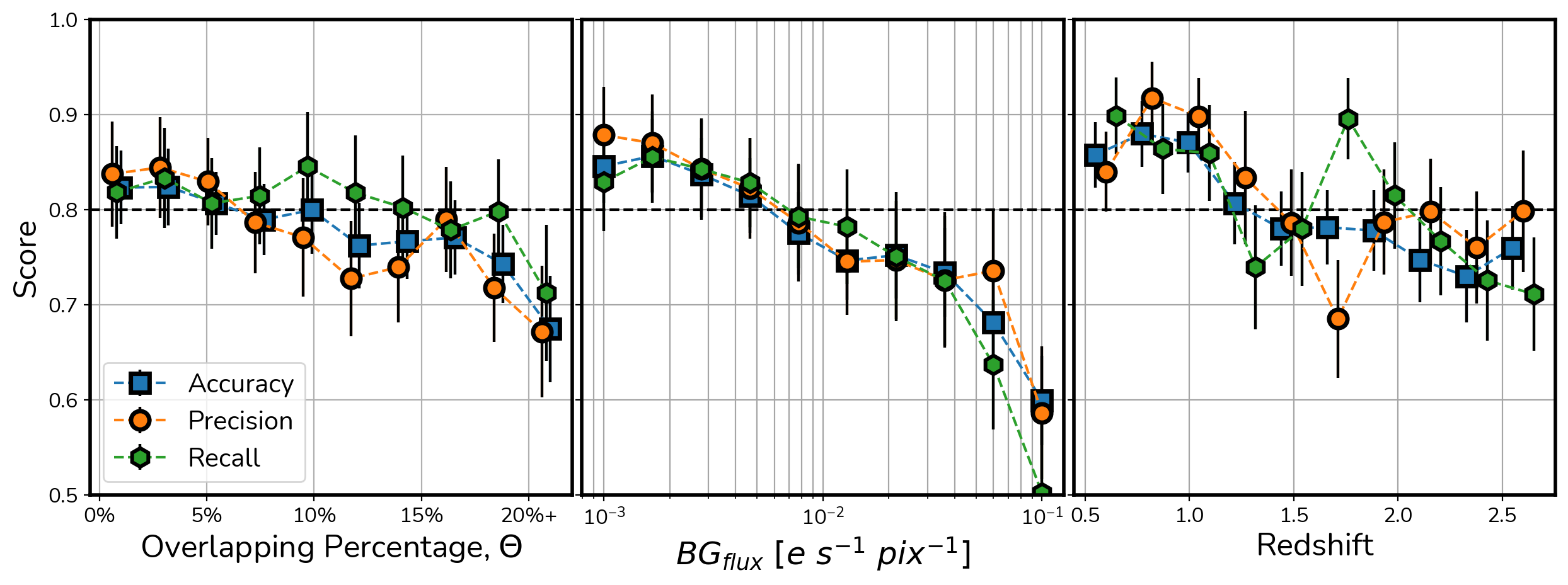}
    \caption{Impact of contamination and redshifts on the performance of our models. The left and central panels show how accuracy, precision and recall (blue squares, orange circles and green hexagons, respectively) behave for increasing percentages of overlap ($\Theta$) and for increased background flux ($BG_{flux}$). In the right panel we show how the accuracy, precision and recall of our methods change in bins of $\Delta z = 0.25$ redshift. Errorbars are sampled from bootstraping the test sample. The performance gradually decreases with $z$, decreasing below $80\%$ beyond $z=2$. There is a slight uptick at $z=2.5$, but with large errorbars. The cutoff at $z>2.5$ is the result of a combination of small sample size and redshift effects. The black dashed line at $score=0.8$ indicates the overall accuracy of the model in the complete test set.}
    \label{fig:performance}
\end{figure*}

With the goal to apply our models to a wide range of redshifts, we explore how our performance metrics are impacted by increasing redshifts. Following the angular size -- distance-relation, galaxies at increasingly larger distances from low to intermediate redshifts will be greatly impacted by decreasing resolution, which means that morphological features are less well sampled. The right panel of Figure \ref{fig:performance} shows this effect on the performance of our models, where the scores of the metrics gradually decrease with increasing redshift, going from $85\%$ accuracy at $z=0.5$ to around $80\%$ at $z=2$. The errorbars -- sampled from boostraping our testing samples -- follow accordingly.

\subsubsection{Contamination impact on classification}

We use the contamination estimates measured in \S \ref{subsub:contquant} to find the contamination failure threshold of our classifier, comparing performance metrics for subsets of the test set selected in bins of both the overlapping percentage, $\Theta$, and the average background flux per pixel, $BG_{flux}$, as shown in Fig.\ref{fig:performance}. The horizontal black dashed line at 0.8 shows the accuracy of the model when evaluated in the complete test set ($80\%$). The metrics outperform this baseline in sub-samples of images with low contamination, decreasing as we increase each of the contamination factors. 

As described in Section \S \ref{subsub:contquant}, we select the point where the average mean values for each metric falls below the dashed line, which is our contamination cutoff, i.e., 
\begin{equation*}
    \Theta \sim 15\%, \quad BG_{flux} \sim 10^{-1.5} \ e \ s^{-1} \ \rm pix^{-1}.     
\end{equation*}
Since it is not possible to directly measure the contamination parameters in the real observations, we refer the reader to our deep learning model trained to measure the contamination in Appendix \S \ref{appendix:contamination}.

\subsection{Classifications on CANDELS}
\label{sec:classifications_on_Candels}

\begin{figure*}
    \includegraphics[width=\textwidth]{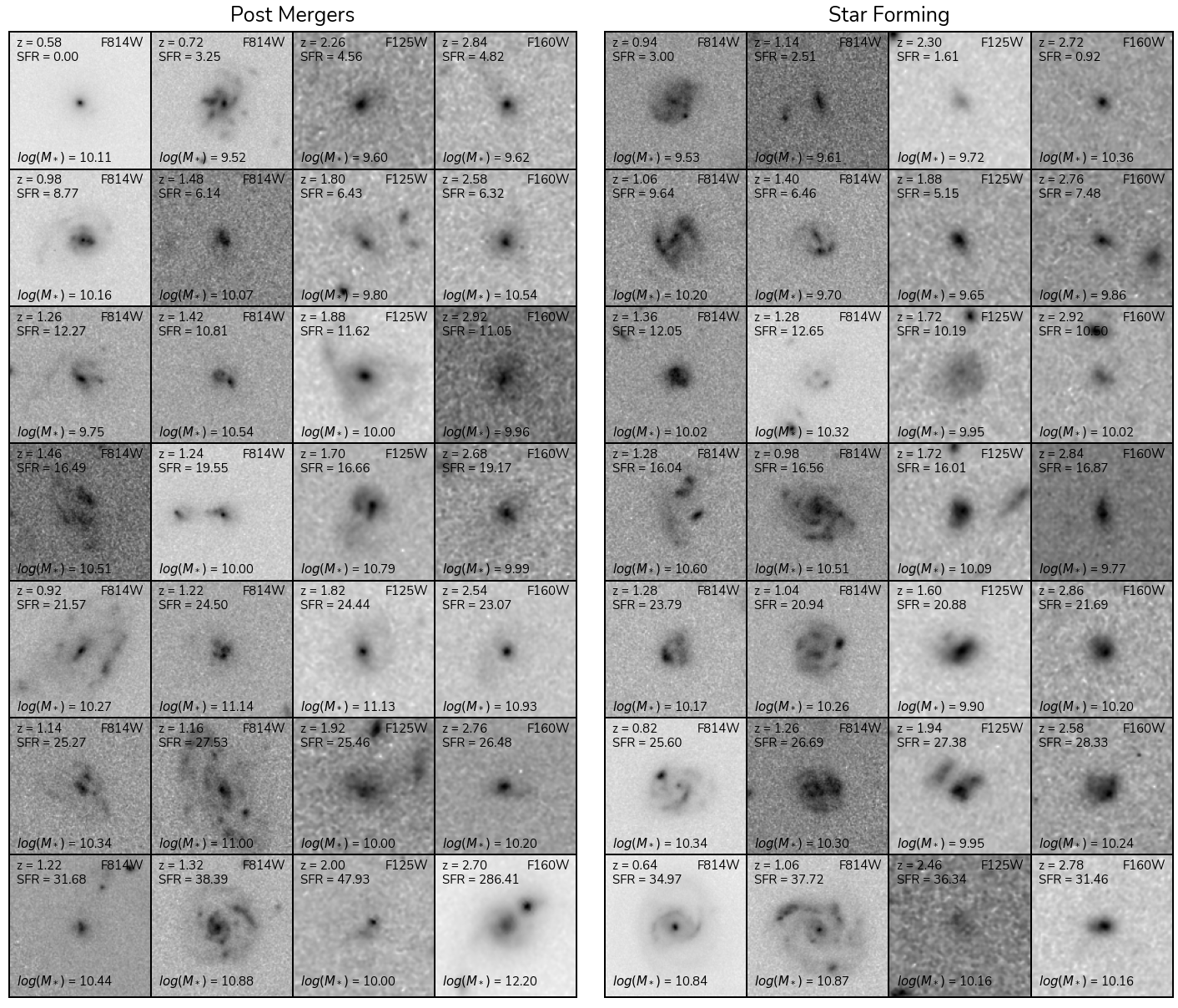}
    \caption{Examples of CANDELS galaxies with $A > 0.1$ classified by our models into Post-mergers (left) and star forming galaxies (right), with their redshifts, SFRs, and stellar masses. Images are ranked from left to right with increasing redshift and top to bottom with increasing SFR. All stamps use a square-root normalization.}
    \label{fig:mosaic_candels}
\end{figure*}

We use our network to carry out predictions in all real CANDELS galaxies at $0.5 < z < 3$, $M_* \ge 10^{9.5} M_{\odot}$, $S/N > 50$ and $H_{\rm MAG} < 24.5$. We filter out regular galaxies using a conservative asymmetry cut of $A > 0.1$ as we are interested only in asymmetric, irregular/peculiar systems. This selection results in a sample of $23,494$ galaxies, for which $14,410$ have visual classifications from \cite{Kartaltepe2015}. Based on the classifications from our networks, we separate these galaxies in post-mergers and non-interacting star forming galaxies using a threshold probability of $60\%$. Galaxies with probabilities $50\% < P(PM) \wedge P(SF) < 60\%$ are not considered in any class. These represent 2125 galaxies ($\approx 15\%$) of the sample with visual classifications. Figure \ref{fig:mosaic_candels} showcases some examples of galaxies in the CANDELS fields separated by the classification of our models. Post-mergers in the left panel and star forming galaxies in the right panel.

\begin{figure*}
    \includegraphics[width=1\textwidth]{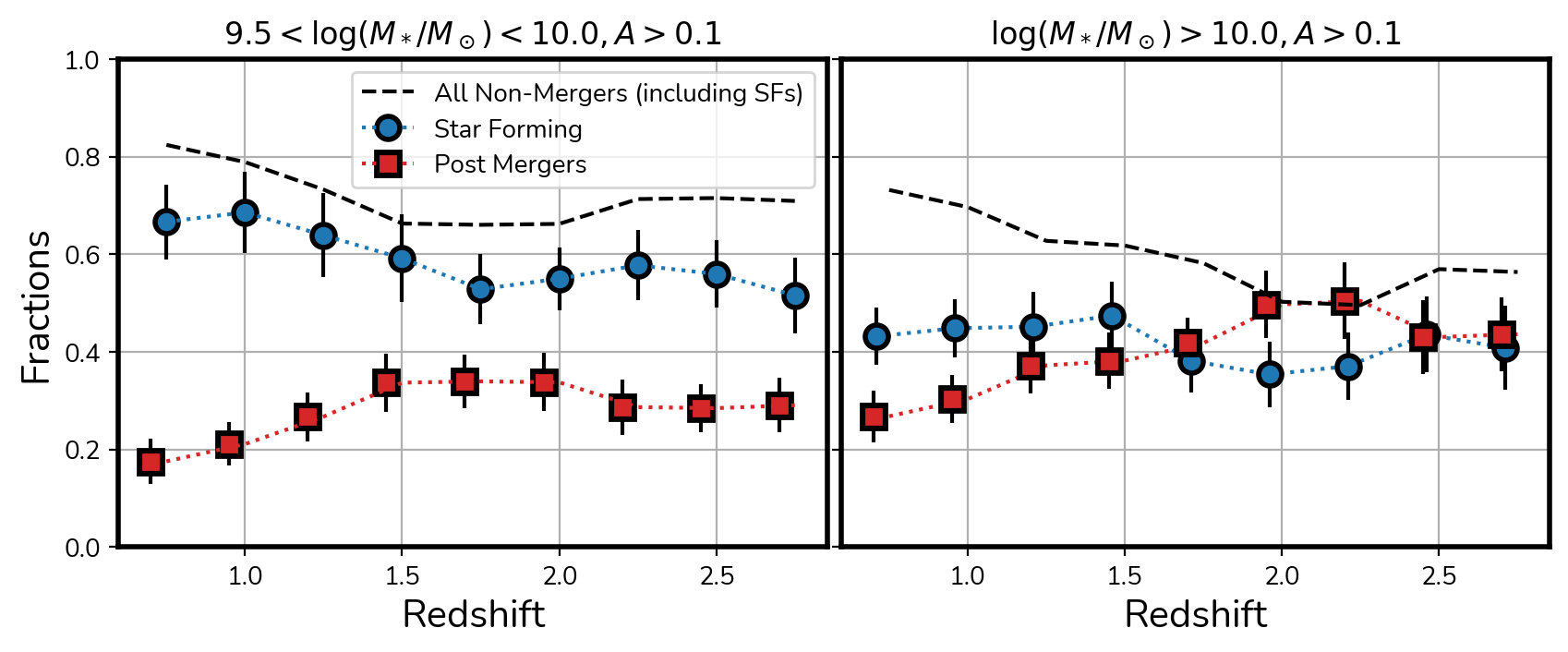}
    \caption{Relative class fractions for post-mergers and star forming galaxies vs. redshift for real galaxies in the CANDELS fields. The fraction of post-mergers increases from 30\% at $z\sim 0.75$ to 50\% by $z \sim 2$. Errorbars are drawn from bootstrapping the samples and applying the underlying uncertainty associated with the performance of our models, which decreases with redshift. }
    \label{fig:starburstiness_with_z}
\end{figure*}

To investigate how the relative number of post-mergers and star forming galaxies changes over cosmic time, we divide the CANDELS sample in bins of $\Delta z = 0.25$. Figure \ref{fig:starburstiness_with_z} shows the change of class fractions change with redshift. We do this analysis in two mass regimes: low mass galaxies with $9.5 < \log(M_*/M_\odot) < 10.0$ (left panel) and high mass systems with $\log(M_*/M_\odot) > 10.0$ (right panel). Non-interacting star forming galaxies are shown in blue circles while post-mergers are shown in red squares. The upper dashed line displays the fraction of galaxies that are not mergers, including the star forming galaxies and other low probability cases not included in any class.

For the lower mass post-mergers we see an upward trend from $\sim15\%$ at $z=0.5$ to $\sim35\%$ at $z=2$, then a slight decrease beyond $z=2$. This is still consistent with a $\sim35\%$ fraction within the error-bars. The star forming galaxies behave in the opposite way, decreasing from $\sim70\%$ at $z=0.5$ to around $\sim55\%$ at $z=2$. This suggests that among asymmetric galaxies of this mass range, there is an exchange between the classes as we go to higher redshifts up to $z=2$. This once again emphasizes that classifications of local galaxies that are purely based on the asymmetry ($A$) are highly contaminated with non-interacting star forming galaxies. However, this is mitigated at higher redshifts where we find more post-mergers. Nevertheless, samples selected based on $A$ are still dominated by star forming galaxies, albeit to a lesser extent.

Trends for higher mass galaxies are substantially different (right panel of Figure \ref{fig:starburstiness_with_z}). While the post-mergers exhibit a similar but more steep upward trend from $\sim20\%$ at $z=0.5$ fraction to $\sim50\%$ at $z=2$, the relative fraction of star forming galaxies show a constant value of $\sim50\%$ at $0.5 < z < 3$, while the fraction of the rest of the sample (dashed line) goes from $\sim75\%$ to $\sim50\%$. This supports the idea that for more massive systems, post-mergers at higher redshifts will eventually become massive passive galaxies with no significant asymmetric features.

At the highest redshift bins, the error bars are large, indication for our networks to perform less accurately above $z=2$. The fraction of post-mergers changes from $30\%$ to around $50\%$ at $z=2$.  We therefore attribute the downtrend in post-mergers beyond $z=2$ to the poor performance of our models at high redshifts and do not take this to imply a real evolutionary effect. 

We know that mergers are more common in the past \citep[e.g,][]{Mundy2017, Duncan2019, FERREIRA2020, Whitney2021}, and here we find further evidence that this is also the case for peculiar galaxies, indicating that the nature behind these disturbed morphologies at earlier times can be attributed to merging. To further investigate this, we select all galaxies from \cite{Kartaltepe2015} that are classified as an irregular / peculiar with {\tt f\_Irr }$ > 0.75$, i.e.,  cases where more than $50\%$ of the visual classifiers agree on the classification, and check how our networks perform on this subset. We observe similar trends with redshift, with the fraction of post-mergers increasing by $\sim 20\%$  from $z=0.5$ ($\sim30\%$) to $z=2$ ($\sim 50\%$), which agree with the results for the complete sample. Furthermore, our methods classify $\sim 50\%$ of the galaxies visually classified as potential mergers ({\tt f\_merger} $> 0.75$) in \cite{Kartaltepe2015} as post-mergers.  This is higher than random, but does show the difficulty of obtaining exact matches between mergers determined visually compared with a quantitative process.

\subsubsection{Visual representation of the classification}\label{subsec:pred_on_candels}

\begin{figure*}
    \includegraphics[width=\textwidth]{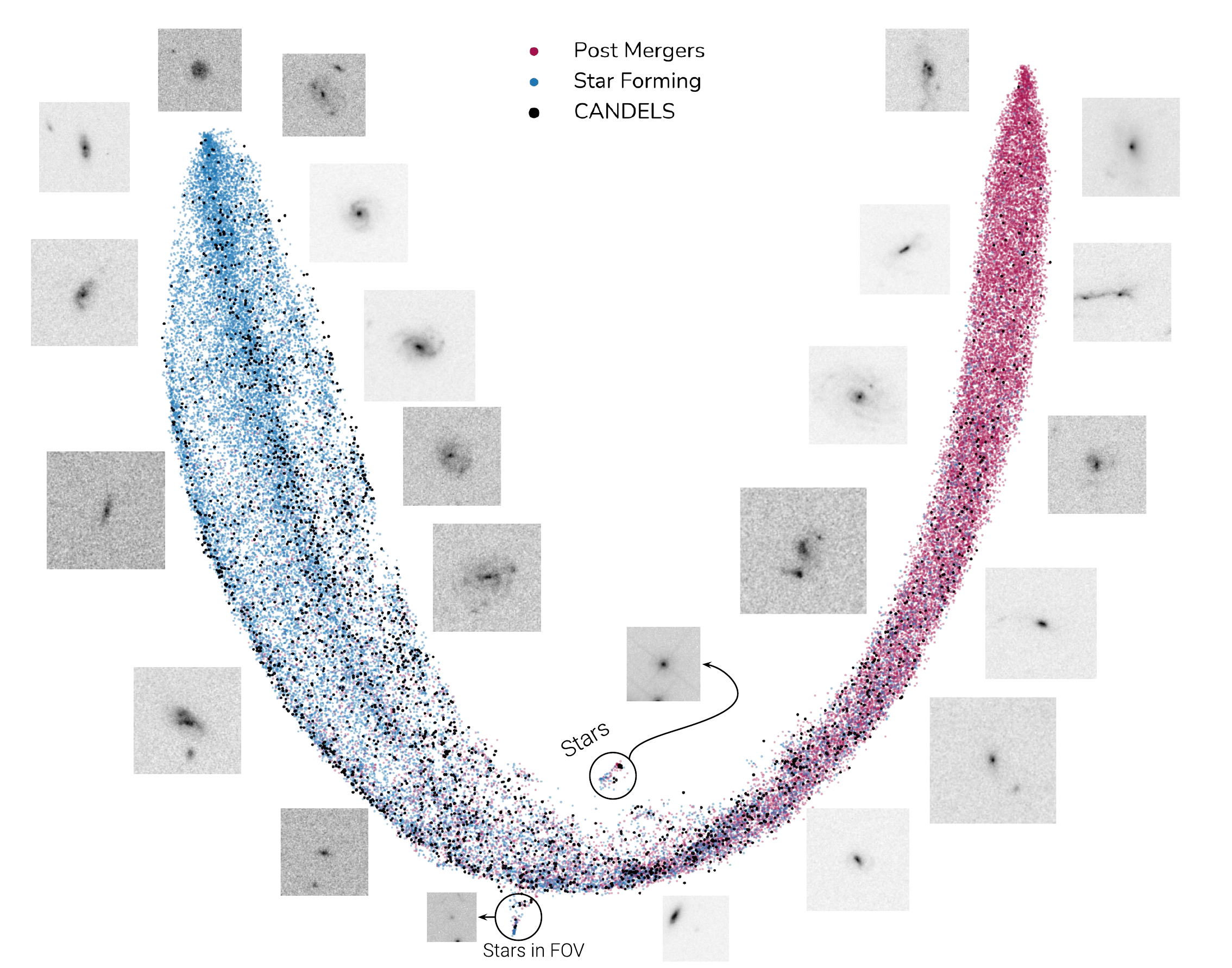}
    \caption{UMAP representation of the output from the last dense layer of the network. This representation shows the parameter space used for the network to generate the final probability. Probabilities are highest in the extremes at the top, and uncertainty increases due to increased contamination as we go along this structure towards the middle. Same random examples of CANDELS galaxies are placed close to their points in this manifold. Small regions identified by circles show the clustering of non-galactic detections in this parameter space, located close to the region of uncertain classifications at the bottom of the UMAP. "Stars" are stars in the center and "Stars in FOV" correspond to stars at the edge of the stamps.}
    \label{fig:decision_boundary}
\end{figure*}

As a way to visualize how our networks organize the features extracted from the images to produce the final classification, we generate a 2D representation of the final dense layer of the network corresponding to 128 neurons (128 dimensions) using an UMAP (Fig. \ref{fig:decision_boundary}). The color code of the points expresses their respective labels, red for post-mergers and blue for star forming galaxies. Then, we overplot the positions assigned by the network for unlabeled CANDELS galaxies. We also include some examples of images of CANDELS galaxies close to their original position in this manifold as a way to visualize how the morphologies change with its position.
Each region in the parameter space of this diagram is directly related to a probability. The maximum probability is found in the extreme regions further away from the center, which represents how different these objects are for the network. Images of galaxies the network struggles to identify are mixed in the bottom middle, representing the region where both probabilities are similar $P(PM) \sim P(SF)$.

\section{Implications}\label{sec:discussion}

\begin{figure*}
    \centering
    \includegraphics[width=\textwidth]{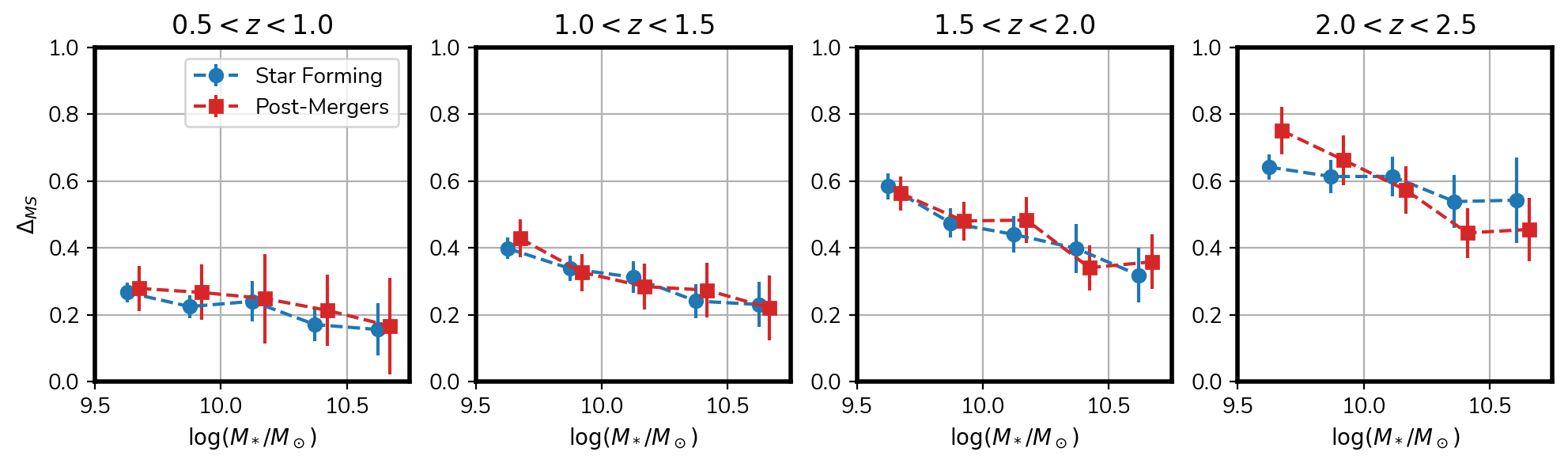}
    \caption{Mean distance to the star formation main sequence ($\Delta_{MS}$) vs. log stellar mass in bins of redshift for CANDELS galaxies above the star formation main sequence. Post-mergers are plotted as red squares, star forming galaxies as blue circles. Errorbars are estimated using bootstrapping and show $\pm 1 \ \sigma$. The classes are indistinguishable between $0.5 < z < 2$, both increase similarly in $\Delta_{MS}$ as we increase redshift. This represents the increase in scatter above the main sequence. All galaxies included in this diagram lie above the SFMS, as we are only interested in exploring the scattering above the SFMS. For the last redshift bin ($2 < z < 2.5$), there is a significant difference between the two classes both at the low mass end and at the high mass end. At low masses ($\log(M_*/M_\odot) < 10.0$), post-mergers scatter higher than star forming galaxies with a difference of $\Delta_{MS} \sim 0.1 \ \rm dex$. At high masses ( $\log(M_*/M_\odot) > 10.0$) the trend reverses and star forming galaxies scatter higher with a $\Delta_{MS}$ difference of $\sim 0.1 \ \rm dex$. 
    However, care needs to be taken in the interpretation of this trend as it could be spurious or insignificant given the errorbars and the performance metrics of our models at high redshift (Fig. \ref{fig:performance}).  
    }
    \label{fig:aboveSFMS}
\end{figure*}

Making use of the classifications from our deep learning models, we first explore the impact of major-mergers on classifications above the Star Forming Main Sequence as parametrized by \cite{Schreiber2015} (\S \ref{sec:above_main_sequence}). We then discuss the structure of the two galaxy classes using S\'ersic profile measurements (\S \ref{sec:structure_implications}). In \S \ref{sec:merger_fractions}, we update classifications from \cite{FERREIRA2020} with our new specialized model, thus increasing certainty for previously undefined classifications. We then add to the discussion proposed by \cite{Bickley2021} regarding the Bayesian limitations of classifying post-mergers by considering an evolving merger fraction. We finish with \S \ref{sec:transfer}, in which we compare extracted features from real CANDELS galaxies to features extracted from IllustrisTNG galaxies, as a way to address the challenges of transferring the model from simulations to real observations.

\subsection{Classifications above the star forming main sequence}\label{sec:above_main_sequence}

The influence of merging on the structure of peculiar / irregular galaxies at intermediate redshifts ($0.5 < z < 3.0$) is directly related to the question of whether merging galaxies can induce more starbursting episodes than galaxies evolving secularly. Enhanced star formation can then lead to more clumpy and asymmetric structures, and thus can impact the morphological appearance of galaxies greatly. By examining the star formation main sequence of galaxies, one can investigate the nature of galaxies with unusually high SFRs and the formation path that resulted in this physical effect.

In order to investigate this, we select only galaxies in our CANDELS fields sample that lie above the star forming main sequence as parametrized by \cite{Schreiber2015}. We separate these sources by stellar masses, redshifts, and their post-merger/star forming classification, measuring the mean distance to the star formation main sequence ($\Delta_{MS}$), as:
\begin{equation}
    \Delta_{MS} = \log(\rm SFR) - \log(\rm SFR_{MS}),
\end{equation}

\noindent where $\log(\rm SFR)$ is the log star formation rate of a particular galaxy and $\log(\rm SFR_{MS})$ is the parametrization from \cite{Schreiber2015}. The SFRs and stellar masses used here for CANDELS galaxies were compiled by \cite{Duncan2019} through SED fitting. We refer the reader to this publication for further details. In Figure \ref{fig:aboveSFMS} we show the mean value of each stellar mass bin, for four redshift ranges (one in each panel), separated into star forming and post-merger galaxies by our classifications. For the $0.5 < z < 2.5$ redshift range (panels A, B and C), we do not find any impactful difference between the classes and  $\Delta_{MS}$, with all offsets well within the errorbars. However, for redshifts $2.5 < z < 3.0$ post-mergers with $\log(M_*/M_\odot) < 10.0$ are on average $\sim 0.1 \ \rm dex$ higher than star forming galaxies of the same mass. The opposite is found for $\log(M_*/M_\odot) > 10.0$, however uncertainty is higher here. Additionally, $\Delta_{MS}$ increases with redshift in all cases, which describes a larger scatter above the star forming main sequence. However, given the performance metrics of our models at high redshift (Figure \ref{fig:performance}), we cannot claim that this is a real effect. We stress that in Figure \ref{fig:aboveSFMS} we only select galaxies above the SFMS, which is why the distance is always positive.

In summary, locations of post-mergers and non-interacting galaxies in the star forming main sequence diagram are comparable, with the possible exception at the highest redshifts. This suggests one of the following: within our sample of CANDELS galaxies, major-merging is not playing a major role in enhancing starbursting episodes; or the timescale probed by our method is too large and the SFR enhancement from the captured post-mergers is short lived.

A relevant result was discussed in \cite{Hani2020}, who investigated TNG300-1 post-mergers at $0.0 < z < 1.0$. They showed that post-mergers have enhanced specific star formation rates by a factor of $\sim 2$, but that this effect decays in timescales of $\sim 0.5 \ \rm Gyr$, which can be driven in part by minor-mergers. Although we do not find evidence for an enhancement in starbursts due to major mergers, we do not rule out the importance of minor mergers to this effect. We trained our models without the presence of minor mergers, but we can not be sure that the star forming galaxies classified by our models are not in some cases triggered by minor mergers.

\subsection{Structure and light profiles}\label{sec:structure_implications}

Our deep learning classifications relate to two different formation pathways. These formation scenarios could result in structures that differ for post-mergers and star forming galaxies. To verify if in fact their structures are diverse from one another, we investigate light profile fitting by using S\'ersic profiles measured by \textsc{Morfometryka}.

\begin{figure}
    \includegraphics[width=0.45\textwidth]{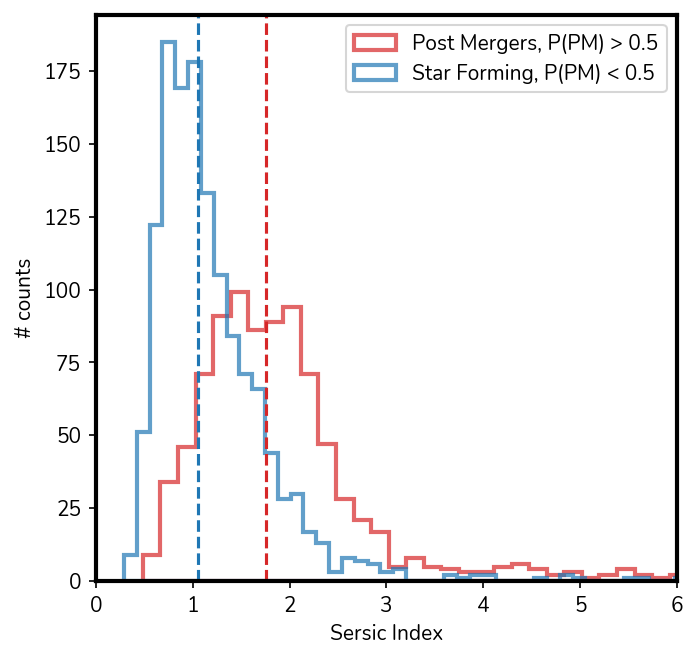}
    \caption{Sersic index distribution for post-mergers and star forming galaxies, in red and blue respectively. Post-mergers display more concentrated light distributions with $n \sim 1.8_{-0.6}^{+0.7}$ while the star forming galaxies have $n \sim 1.1_{-0.5}^{+0.5}$ consistent with disk dominated galaxies.}
    \label{img:sersic_balanced}
\end{figure}

Figure \ref{img:sersic_balanced} shows the distribution of S\'ersic indices for post-mergers in red, and star forming galaxies in blue. In general, each class presents very distinct distributions: the post-mergers have a mean S\'ersic index $n \sim 1.8_{-0.6}^{+0.7}$ roughly representative of a transition from disks to spheroids;  star forming galaxies have systematically lower S\'ersic indexes with $n \sim 1.1_{-0.5}^{+0.5}$, which is more consistent with disk dominated galaxies. This offset of $\sim 1 \ \rm dex$ increases for classification thresholds at higher values. The average S\'ersic profile ($n$) of post-mergers increases while the distribution for star forming galaxies continues with a similar shape. This is quantitative evidence that 1) post-mergers with higher light concentrations are more easily separable from non-interacting star forming galaxies, and 2) these types of galaxies are intrinsically different from each other.

\subsection{Merger Fractions and Rates}\label{sec:merger_fractions}

By using the new classifications from this work we can update classifications from \cite{FERREIRA2020} for cases where the previous method had ambiguous probabilities for some major-mergers and non-mergers. 

\begin{figure}
    \centering
    \includegraphics[width=0.5\textwidth]{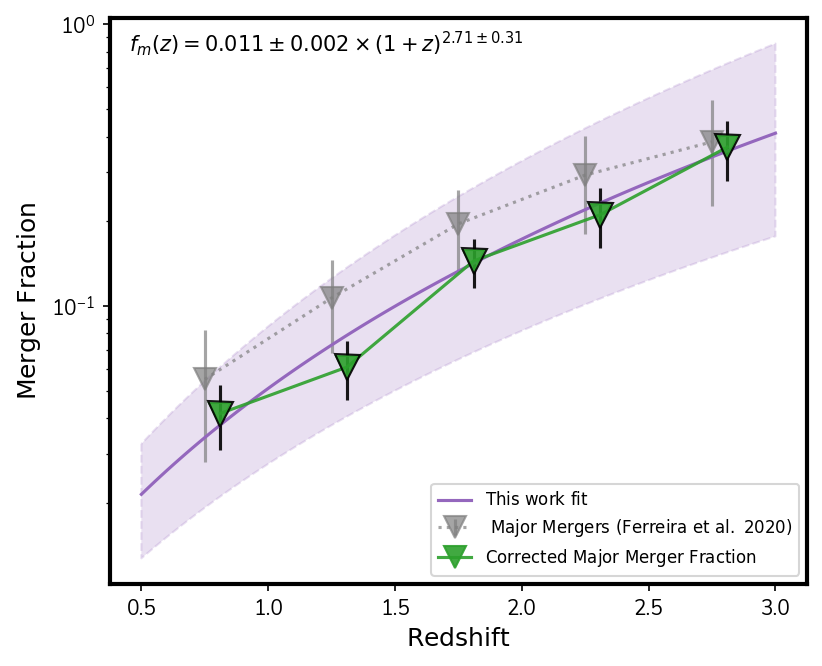}
    \caption{Major-merger fractions as a function of redshift. We show corrected merger fractions from \cite{FERREIRA2020} by re-classifying galaxies with our new method in mergers and non-mergers, shown in green. The original estimates are shown in gray.}
    \label{fig:merger_history}
\end{figure}

Our new dataset accounts for the effects of dust, it is not limited by orientation and probes the rest-frame optical. Thus we can check if any major merger classifications in the previous work can be attributed to non-interacting star forming galaxies or if any non-mergers can be re-classified as post-mergers. This is done by comparing the probabilities for major-mergers and non-mergers, $P(MM)$ and $P(NM)$, respectively, from \cite{FERREIRA2020} to the new probabilities $P(PM)$ and $P(SF)$. We update a non-merger classification to post-merger if
\begin{equation*}
    P(PM) > P(NM),
\end{equation*}
and update the major-merger classifications to non-merger if
\begin{equation*}
    P(SF) > P(MM).
\end{equation*}
In other words, we reclassify galaxies from the previous sample where our new method is more certain about its classification than the previous one. This leads to $\sim 5\%$ of major-mergers reclassified as star forming non mergers, which lowers the overall merger fractions at lower redshifts and keeps it similar at higher redshifts. In Figure \ref{fig:merger_history} we compare the new merger fraction measurements, in green, to the results from \cite{FERREIRA2020}, in gray. 

The updated fit of the cosmic evolution of the merger fraction, $f_m(z)$
\begin{equation}
    f_m(z) = 0.011 \pm 0.002 \times (1+z)^{2.71\pm0.31},
\end{equation}
with errors estimated with boostraping, agrees with the previous measurement in \cite{FERREIRA2020} within errors. To measure the galaxy major-merger rate ($\mathcal{R}$), we combine the timescale ($\tau_{m} = 0.5 \ \rm Gyr$) used in our selection (\S \ref{sec:illustris_data}) with this merger fraction through
\begin{equation}
    \mathcal{R} = \frac{f_m}{\tau_m}.
\end{equation}
The updated galaxy major merger rate is
\begin{equation}
    \mathcal{R} = 0.022 \pm 0.006 \times (1+z)^{2.71\pm0.31}.
\end{equation}
We emphasize that this correction is a minor adjustment to the galaxy major-merger rates presented in \cite{FERREIRA2020}, which remain broadly consistent with each other.

\subsection{Bayesian Analysis of Mergers}
\begin{figure*}
    \includegraphics[width=\textwidth]{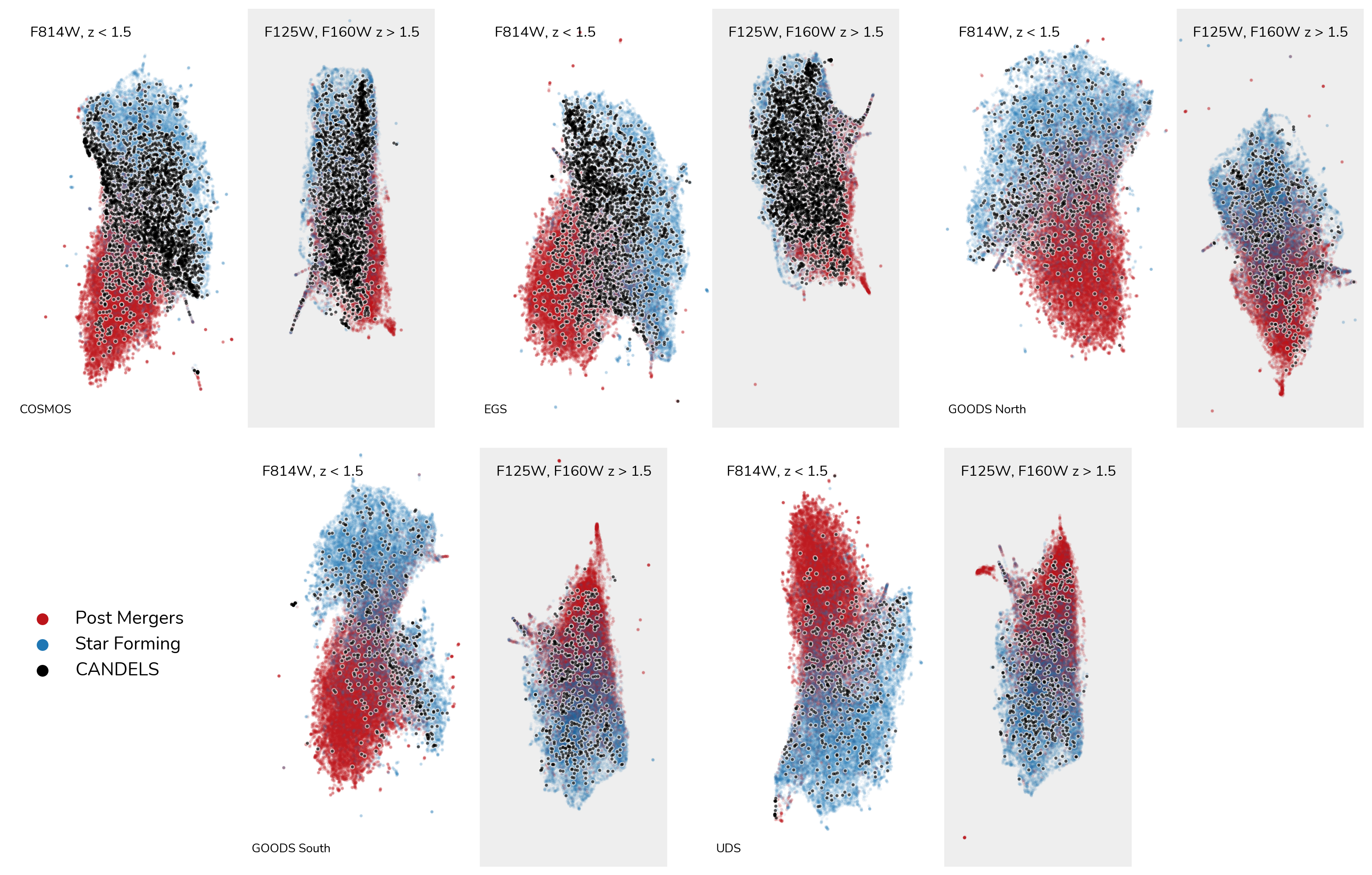}
    \caption{Extracted features by our networks in a UMAP 2D representation. For each model in our ensamble we generate a UMAP from the extracted features from the last convolutional layer of the trained networks, both when applied to the Illustris galaxies, color-coded by the class, and to the unlabeled CANDELS galaxies shown in black dots. Both Illustris and CANDELS extracted features populate the same region of this representation, showing that the features used by the network to then perform the classification task are in general domain invariant. Additionally, both classes -- post-mergers and star forming galaxies -- form separated clusters with some overlapping. Classification could be done in this representation alone, but it is then better organized by the fully connected layers that combine these features to produce the final output probability.}
    \label{fig:domain_adaption}
\end{figure*}


We now investigate the possible contamination in merger samples that are selected through our method. This approach is fairly direct and based on Bayesian statistics, and relies on some understanding of the true intrinsic merger fraction and how it evolves with time.  It also requires that we have a good understanding of the fraction of contamination in merger samples \citep{Bickley2021}.  The basic Bayesian formula to understand this is given by the following.

\begin{equation}
    P(\rm \mathbf{M|S}) = P( \mathbf{S|M}) \times \frac{P \left ( \rm \mathbf{M}\right)}{P \left(\rm \mathbf{S}\right)}, 
    \label{eq:bayes}
\end{equation}

\noindent where $P(\rm \mathbf{M|S})$ is the probability of a merger, given that a method used to select mergers, ($\rm \mathbf{S}$),  identifies it as such.  The value of P($\rm \mathbf{M}$) is the probability that an object is a merger before a selection of merger is made.  P($\rm \mathbf{S}$) is the probability that a galaxy is selected as a merger, whether a real merger or a false-positive. Because of the results of this paper, we know that this last number is very likely not equal to unity.  It in fact can depend on various factors and methods of finding mergers.    We can write the probability P($\rm \mathbf{S}$) as:

\begin{equation}
     P \left(\rm \mathbf{S}\right) = P(\rm \mathbf{S|M}) \times P(\rm \mathbf{M}) + P(\rm \mathbf{S|NM}) \times P(\rm \mathbf{NM})
\end{equation}

\noindent where $\rm \mathbf{NM}$ standard for non-mergers, where P$(\rm \mathbf{S|NM})$ is the probability of identifying correctly a non-mergers, and the value of P($\rm \mathbf{NM}$) is the probability that the galaxy is not a merger.  We can simplify this if we know, a priori, what the merger fraction is based on previous work.  If we denote the merger fraction as $f_{\rm m}$, and the machine learning probability of finding a merger/non-merger as $p_{\rm m}$ and $p_{\rm mn}$ then we can rewrite equation (\ref{eq:bayes}), as:

\begin{equation}
    P(\rm \mathbf{M|S}) =  \frac{p_{\rm m} \times f_{\rm m}}{ (p_{\rm m}  f_{\rm m} + (1-p_{\rm nm}) (1-f_{m})) }. 
\end{equation}

\noindent Thus, for example, if the accuracy of a machine learning method for finding a merger is 0.9 and the accuracy for finding a non-merger is 0.9, and the merger fraction $f_{\rm m}$ = 0.1, then the probability that a galaxy identified as a merger is actually a merger is $P(\rm \mathbf{M|S}) = 0.5$.  This implies that even when the accuracy of finding mergers and non-mergers is $90\%$, at the lowest redshifts, where the merger fraction is low $\sim 10\%$, there is still a 60\% chance that an identified merger is identified incorrectly as such. At higher redshifts, where the intrinsic merger fraction is higher, the probability of finding a merger correctly increases to $\sim 80\%$ when the merger fraction is as high as $f_{\rm m} \sim 0.3$.  

We can generalise the equation for $P(\rm \mathbf{M|S})$, as a function of $z$, by considering how the merger fraction $f_{\rm m}$ evolves with redshift, such that:

\begin{equation*}
    P(\rm \mathbf{M|S}, z) =  \frac{p_{\rm m}(z) \times f_m(z)}{ p_{\rm m}(z)  f_m(z)  + (1-p_{\rm nm}(z)) (1-f_m(z) )}, 
\end{equation*}
\begin{equation*}
    f_m(z)  \equiv f_0 (1+z)^m,
\end{equation*}
\noindent which gives us a tool to understand how our classifications might be contaminated by sample unbalance effects with respect to redshift.

From this we can conclude that a significant fraction of individual galaxies within the CANDELS imaging may be incorrectly identified as either mergers or non-mergers. From our results here, our method effectiveness for correctly classifying mergers increases from $\sim 40\%$ at $z\sim0.5$ to $\sim70\%$ at $z \sim 3$. This is likely what can account for some of our misidentified galaxies as discussed in \S \ref{sec:results} when discussing the success of our method of separating star forming systems from those that are undergoing mergers.
  
These are conservative estimates which do not include the fact that we pre-select CANDELS galaxies based on their asymmetry. This should increase $P(\rm \mathbf{M|S}, z)$ further since $f_{\rm m}$ is higher among galaxies with $A>0.1$.

\subsection{On Domain Adaptation Issues}\label{sec:transfer}

There is a growing concern on the applicability of simulation trained deep learning models when applied to a intrinsically different domain. For us this is the case with going from cosmological simulations to real observations. When transferring from one domain to another, deep learning models might fail due to relying too much on domain specific features. Several techniques were developed to address this problem, focused on forcing neural networks to learn domain invariant features, leading to more robust models. \cite{deepMergeII} show that adopting techniques for domain adaptation could increase model performance when applying to the target domain by $20\%$.

In our case, the source domain are the IllustrisTNG galaxies and the target domain the CANDELS observations. To check if we need to apply domain adaptation techniques to this particular problem, we used UMAPs (Uniform Manifold Approximation and Projection, as described in Sec. \ref{subsec:pred_on_candels}) \citep{UMAPs} to reduce the high dimensional space generated by the features extracted by our network to a 2D-space that is easy to visualize\footnote{We also tested with t-SNEs with similar results.}. Then, for each of our trained models we compare whether the features extracted by the network show similar distributions for Illustris and CANDELS galaxies. In Figure \ref{fig:domain_adaption} we show UMAPs for each of the CANDELS fields models, for low redshift (left) and high redshift (right), color coded by their class in the case of Illustris and in black for real CANDELS galaxy images. As can be seen, these distributions of simulated galaxies and real observations are clustered together, with very few outliers not following the main cluster. Additionally, we can see that each class forms its own cluster, with overlapping regions, showing that features between classes are distinct and in general not domain specific.\footnote{We also tested generating random noise images to check their position in this parameter space. As expected, they cluster away from the image regions, forming its own outlier region which is far from the main locus where galaxies are found.}

We attribute the generalization success of our models to our mock data pipeline, which is tailored to mimic each individual CANDELS field with maximum fidelity -- with their instrumental and observational features. Also augmentations with patches of the sky from CANDELS introduce real observations into our source domain, which not only make our training sets big ($\sim 140.000$ images) but also help with domain confusion within the network. Thus, we do not include any domain adaptation process in our pipeline.

\section{Summary}\label{sec:summary}

To shed light on the nature of peculiar/irregular objects at intermediate to high redshifts, we have constructed a framework based on forward-modelling of cosmological simulations with deep learning algorithms, that allows classifications with physically motivated labels based on the formation history of galaxies. 

We used data from the IllustrisTNG TNG100-1 simulation to create realistic mocks of galaxies with CANDELS-like properties, including a full radiative transfer treatment with SKIRT for two specific classes of galaxies: post-mergers and non-merging star forming galaxies. These are selected so that their main difference is their formation history. 

We produced a dataset of $\sim 160,000$ images of simulated IllustrisTNG galaxies with realistic visual properties that mimick CANDELS observations in the redshift range $0.5 < z < 3.0$. The images are used to train Deep Convolutional Neural Networks to distinguish between formation histories of post-mergers and star forming galaxies. The main conclusions drawn from this work are summarized as follow:

\begin{itemize}
    \item The classifier network combined with our new dataset produces classification models with a balanced performance of $\sim80\%$ accuracy, precision, and completeness when applied to a single-band imaging  dataset, outperforming the asymmetry ($A$) by at least 25\% within the simulated data. Additionally, for pristine images without any contamination and observational effects, the theoretical limit of our model is $\sim 91\%$ accuracy. This is evidence that using the asymmetry ($A$) alone for ambiguous morphological cases might generate highly contaminated samples.  
    
    \item We define two new contamination indicators, the overlapping percentage, $\Theta$, and the average flux of the background sources, $BG_{flux}$, by leveraging how simulated galaxies are combined with true CANDELS background sky patches. $\Theta$ controls how sources overlap and are projected in the same stamp, while the $BG_{flux}$ value probes the effect of the brightness of external sources on the classification of the central object. These allow us to explore in detail how deep learning classifications are impacted by contamination. We show that both crowded environments and projections and the relative brightness of external sources to the central galaxy negatively impacted our results. Based on this, we define quality control limits to our approach within the CANDELS fields as $\Theta \sim 10\%$ and $BG_{flux} < 10^{-3}$\ e \ s$^{-1}$ \ pix$^{-1}$. Although not universal, these limits provide guidelines for sample selection when applying our models to data. 
    
    \item By applying our model to real CANDELS observations of galaxies with high asymmetries, we show that the relative fraction of post-mergers to star forming galaxies increases with higher redshift for two mass regimes. For low mass sources ($9.5 < \log(M_*/M_\odot) < 10.0$), the post-merger fraction increases by $\sim 20\%$ within $0.5 < z < 2.0$, while the fraction of star forming galaxies decreases by $\sim 15\%$ in the same redshift range. In the high mass case ($\log(M_*/M_\odot) > 10.0$), the post-merger fraction increases by $\sim 25\%$ at $0.5 < z < 2.0$, while the fraction of star forming galaxies stays broadly constant. 
    
    \item We explore the impact of major mergers on galaxies located above the star formation main sequence (SFMS) as parametrized by \cite{Schreiber2015}. We separate CANDELS galaxies above the SFMS in the classes provided by our model and in bins of stellar mass. At $0.5 < z < 2.0$ we do not find any clear signs that major mergers play a critical role on the scattering above SFMS, with similar trends for post-mergers and star forming galaxies. However, in the highest redshift bin with good sample statistics ($2.0 < z < 2.5$) we see a post-merger driven SFR enhancement at lower masses of about $\sim 0.1 \rm \ dex$.
    
    \item We show that the light distribution parametrized through S\'ersic profiles of the CANDELS galaxies classified by our models as post-mergers are intrinsically distinct from those classified as star forming galaxies. The star forming galaxies sample is dominated by disk-like objects with an average S\'ersic index of $n=1.1_{-0.5}^{+0.5}$ while the post-mergers have more concentrated light profiles corresponding to higher central concentration with $n=1.8_{-0.6}^{+0.7}$, with a long tail at higher S\'ersic indices. Moreover, when we increase the probability threshold of our classifications to improve the purity of our selections, only the post-merger distribution display higher S\'ersic indices. Evidently, our model predicts that post-mergers are more likely to be bulge-dominated galaxies.
    
    \item By using our updated data pipeline and models specifically tailored to distinguish between post-mergers and star forming galaxies, we revisit the merger fractions and merger rates from \cite{FERREIRA2020} by correcting ambiguous cases. This leads to updated galaxy merger rates that are slightly lower, but consistent with previously reported rates:
    $\mathcal{R} = 0.022 \pm 0.006 \times (1+z)^{2.71\pm0.31}$. 
    
    \item We show that our models use similar features to classify IllustrisTNG and real CANDELS galaxies, with no clear discrepancy between the two domains. Using the features extracted by the convolutional layers of our network, we generate UMAPs, which visualize the complex parameter space in 2D. Features of IllustrisTNG galaxies and CANDELS galaxies overlap for all the CANDELS fields. Although the CANDELS galaxies do not span the entire feature space of the IllustrisTNG galaxies used here, they are contained within that feature space. 

\end{itemize}

Our machine learning driven approach provides a new way to investigate the formation history of galaxies with models that are informed by cosmological simulations. This includes the use of the models themselves, and the application of these models within accurate observing conditions.

Nevertheless, currently we are still limited to high-mass major merger cases due to resolution limitations from the simulations and mass completeness from the observations. In the upcoming years, combining the next generation of high resolution, small box simulations (e.g. TNG50-1, New Horizons) with observational data from the James Webb Space Telescope (JWST) and \textit{Euclid} Telescope will open a new window to incorporate the effect of minor mergers and lower mass systems. Together, this will represent a major step towards uncovering unresolved questions of galaxy evolution.

\section*{acknowledgments}
The authors thank the anonymous referee for the detailed review that improved the paper greatly.
The authors thank Centre for Astronomy and Particle Theory of University of Nottingham for providing all computational infrastructure
necessary to run the training steps to produce the model
described here. This study was financed in part by the
Coordernação de Aperfeiçoamento de Pessoal de Nível
Superior - Brazil (CAPES). CJC acknowledges support from the European Research Council (ERC) Advanced Investigator Grant
EPOCHS (788113). UK acknowledges support from the Science and Technology Facilities Council through grant number RA27PN. The IllustrisTNG simulations were undertaken with compute time awarded by the Gauss Centre for Supercomputing (GCS) under GCS Large-Scale Projects GCS-ILLU and GCS-DWAR on the GCS share of the supercomputer Hazel Hen at the High Performance Computing Center Stuttgart (HLRS), as well as on the machines of the Max Planck Computing and Data Facility (MPCDF) in Garching, Germany.

\section*{Data Availability}

Ready to use models are publicly available for anyone to download\footnote{https://github.com/astroferreira/FERREIRA2022}, with accompanying code.


The post-processed IllustrisTNG data used for training is stored in large {\tt TFRecords} binary files and are available upon request.

\bibliography{references} 

\appendix

\section{Contamination Network} \label{appendix:contamination}

Based on the contamination measurements described in \S \ref{subsub:contquant}, we devise a new neural network with the goal to predict the overlapping percentage ($\Theta$) and the background flux ($BG_{flux}$) measurements from real observations. In this way, contamination thresholds can be applied to real observational samples in a similar way to what is done in the simulations.

The contamination quantification depends on our ability to separate the background patch of the sky from the central source, a feature that is only available when we are post-processing simulated galaxies. In the case of real CANDELS observations, directly measuring these properties is difficult, because it is not straightforward to de-blend background/foreground sources if they are projected on top of one another or are close enough to be a potential interaction. 

We use all the contamination information from our data pipeline (\S \ref{sec:pipeline_imaging_data}) to train a neural network to predict these values from the final image, without separating source and background. We use the same network architecture described in this work, but replacing the final sigmoid layer with a linear activation function, changing the loss function as well. The result is a model that can be directly applied to real observations, where the image is the input and the outputs are values for $\Theta$ and $BG_{flux}$. 

\begin{figure}[h!]
    \centering
    \includegraphics[width=0.95\textwidth]{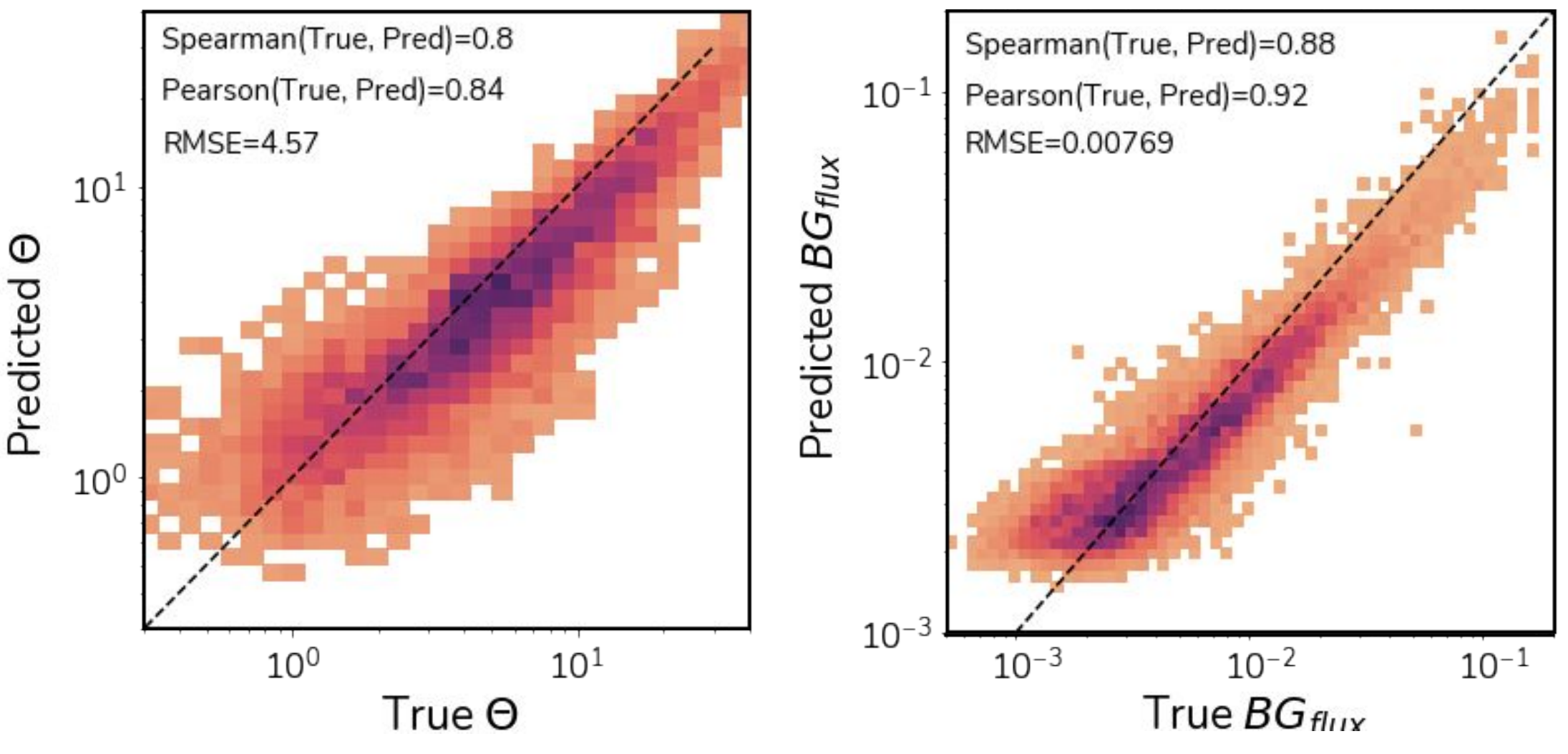}
    \caption{Performance of the contamination quantification network. (top) The relationship between true and predicted values for $BG_{\rm flux}$ and (bottom) relationship between truth and predicted values for $\Theta$. Pearson and Spearman correlation indices are displayed for each case, as well as the root mean square error.}
    \label{fig:appendix_performance}
\end{figure}

\begin{figure*}
    \includegraphics[width=\textwidth]{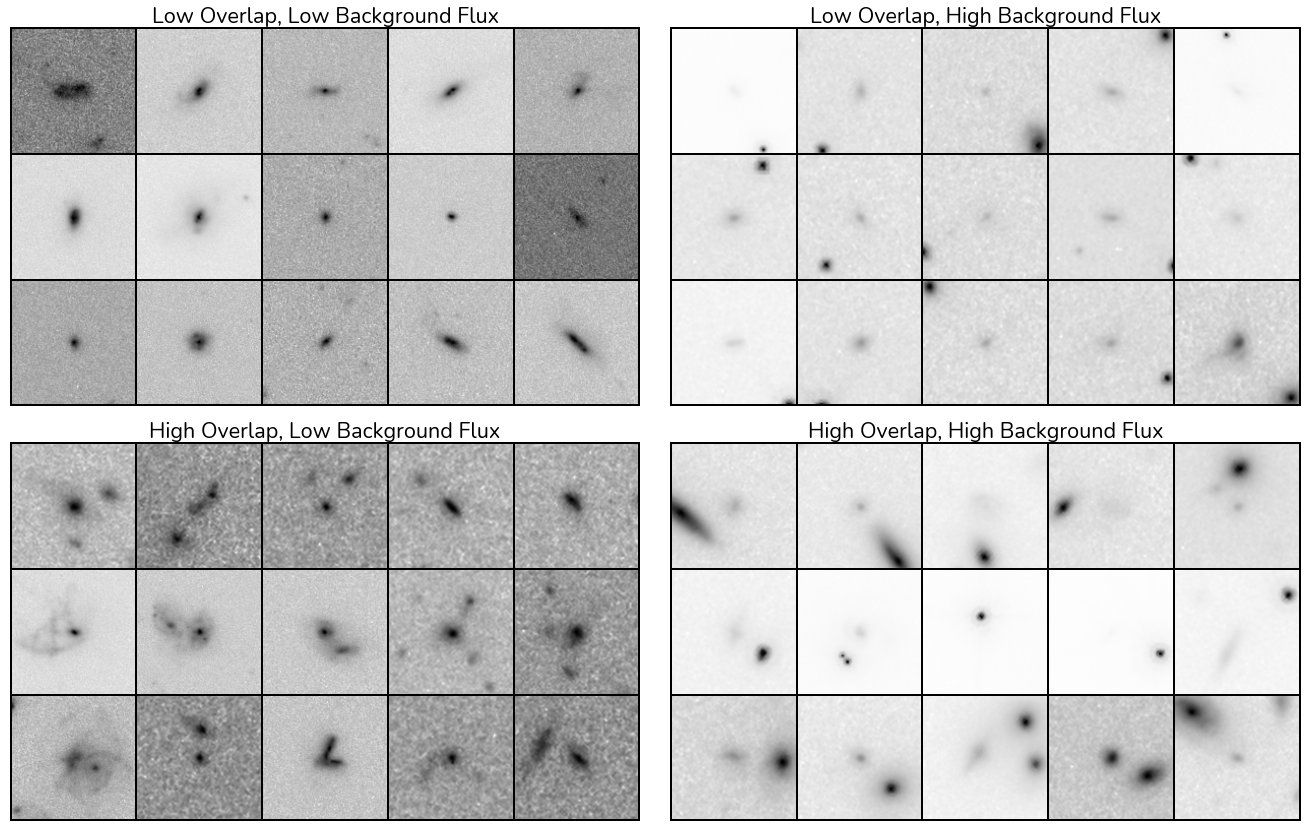}
    \caption{Panel with cutouts of IllustrisTNG galaxies demonstrating four different selections on the $O_f$ and $<f_{BG}>$ parameter space. Isolated galaxies with almost no noticeable contamination have low overlap and low background flux (top left). Low overlap and high background flux show cases where the central galaxy is overshadowed by a bright companion, but with no overlapping (top right). High overlap and low background flux show galaxies overlapping with similar brightness, cases where projection effects can be misinterpreted as a major merger (bottom left). High overlap and high background flux show central galaxies with very large and bright companions that extend over its segmentation map (bottom right). This illustrates how useful these two measurements can be for proper selections.}
    \label{fig:panel_contamination}
\end{figure*}

Figure \ref{fig:appendix_performance} displays the performance of these predictions based on the original measurements, together with Pearson and Spearman correlation indices. In general, the performance of the model is in good agreement with the original measurements, with root mean square errors in the order of $\sim 10^{-3}$ for $BG_{flux}$ and $\sim  5\%$ for $\Theta$. These limits are well within the region of the parameter space formed by these indices that we defined as a low contamination region. Apart from the small bias making the predictions undervalue the truth values, the performance is good enough to separate high contamination cases from the rest of the sample, which is ultimately our goal. In Figure \ref{fig:panel_contamination} we show examples of different combinations of $\Theta$ and $BG_{\rm flux}$.

Even though this network is designed to be used within the context of this work as a way to reproduce contamination quantification in the same manner as what was done with the simulated images, we recognize that this can be useful for a wider application. For example, this can be used as a fast selection tool that can remove catastrophically bad cases from big samples in just a couple of seconds, thus it can be a powerful tool for quick exploration. In this regard, we release this model independent of the classification models presented in \ref{sec:deeplearning}. 

\end{document}